\documentclass[twoside,twocolumn]{article}

\usepackage{blindtext} 
\RequirePackage{amsgen}
\usepackage[sc]{mathpazo} 
\usepackage[T1]{fontenc} 
\usepackage{microtype} 
\usepackage{amsmath}
\usepackage[english]{babel} 
\usepackage{xcolor}
\usepackage[hmarginratio=1:1,top=32mm,columnsep=20pt]{geometry} 
\usepackage[hang, small,labelfont=bf,up,textfont=it,up]{caption} 
\usepackage{booktabs} 

\usepackage{lettrine} 
\usepackage{tabularx}
\usepackage{enumitem} 
\setlist[itemize]{noitemsep} 

\usepackage{abstract} 

\usepackage{titlesec} 
\renewcommand\thesection{\Roman{section}} 
\renewcommand\thesubsection{\roman{subsection}} 
\titleformat{\section}[block]{\large\scshape\centering}{\thesection.}{1em}{} 
\titleformat{\subsection}[block]{\large}{\thesubsection.}{1em}{} 

\usepackage{fancyhdr} 
\usepackage{titling} 

\usepackage{hyperref} 

\pretitle{\begin{center}\Huge\bfseries} 
\posttitle{\end{center}} 
\title{Symplectic Reduction of Classical Mechanics on Shape Space} 
\author{%
\textsc{Sahand Tokasi}\thanks{sahand.tokasi@uni-tuebingen.de} \\[1ex] 
\textsc{Peter Pickl}\thanks{p.pickl@uni-tuebingen.de} \\[1ex]
\normalsize University of T\"ubingen \\ 
}
\date{\today} 
\renewcommand{\maketitlehookd}{%
\begin{abstract}
\noindent 
One of the foremost goals of research in physics is to find the most basic and universal theories that describe our universe. Many theories assume the presence of an absolute space and time in which the physical objects are located and physical processes take place. However, it is more fundamental to understand time as relative to the motion of another object, e.g. the number of swings of a pendulum, and the position of an object primarily as relative to other objects.\newline 
The goals of this paper is to explain, how using the principle of relationalism (to be introduced below), classical mechanics can be formulated on a most elementary space, which is freed from absolute entities: shape space. On shape space only the relative orientation and length of subsystems are taken into account. 
In order to find out how the shape of a classical
system evolves in time, the method of "symplectic reduction of Hamiltonian systems" is extended to include scale
transformations, and in this way the reduction of a classical system with respect to the full similarity group
is achieved. A necessary requirement for the validity of the principle of relationalism is that changing the length scale of a system, all parameters of the theory that depend on the length, get changed accordingly. In particular, the principle of relationalism requires a proper transformation of the coupling constants of the interaction potentials in Classical Physics. This leads consequently to a transformation in Planck's measuring units, which enables us to derive a metric on shape space in a unique way. Later in this paper, we will explain the derivation of the reduced Hamiltonian and symplectic form on shape space.\newline\newline

\centering We dedicate this paper to Prof. Dr. Detlef D\"urr.

\end{abstract}
}
 
\begin{document}

\maketitle 

\tableofcontents

\newpage
\section{Introduction}
This paper can be divided to 3 parts. \newline\newline

In the \textit{first part} (section II), we review the foundations of classical mechanics with an emphasis on the ideas of \emph{Leibniz}, and compare the Leibnizian relational worldview with absolute worldview of Newton. One of the most basic building blocks of Newtonian mechanics is the idea of an absolute time and space, the existence of which were assumed by \emph{Isaac Newton} when formulating the laws of motion. We will take the absoluteness of space and time into question and will review how Newton's absolute time can be deduced from the change in the positions of particles. By this approach, time loses its status as a primitive notion in physics, and takes an emergent status instead. Our dynamics will be defined on shape space $S$ \footnote{quotient of absolute configuration space $Q\cong\mathbb{R}^{3N}$ with respect to the similarity group $sim(3)$, which  comprises all global spatial translations, rotations, and scalings.}, for which, in contrast to the configuration space used in Newtonian Mechanics, time and space are not absolute entities. The central new concept in the discussion is  the \textit{principle of relationalism}, which forces us to consider particular constants of nature as homogeneous functions of proper degrees on the universe's configuration space.    \newline\newline

 In the \textit{second part} (sections III and IV) of this paper we give a review of the literature on symplectic reduction of classical systems with respect to the Euclidean group $E(3)$. Here we follow \cite{5} and \cite{6} to a big extend. \newline \newline
 
In the \textit{third part} of this paper (section V), we explain how these methods can be expanded to include scale transformations, and consequently how the reduction of a classical system with respect to the full similarity group $sim(3)$, which we consider in this manuscript to be the symmetry group, is achieved. Here we explain among others, how the kinetic metric of configuration space leads in a unique way to a metric on shape space, using the principle of relationalism. Given the known procedures of deriving the reduced equations of motion with respect to the Euclidean group $E(3)$ in the Hamiltonain formalism using a symplectic structure on phase space, we derive the reduced symplectic form, and the Hamiltonian of a $N$ particle system on its reduced phase space with respect to the similarity group. This suffices to determine the evolution of a classical system on shape space, given its initial shape and shape velocities, without any reference to the system's orientation, position, or scale in absolute space.  \newline

\newpage
\section{Relationalism}
\subsection{Shape Space as the Physical configuration space}

The goal of physics is to give a most accurate description, prediction and understanding of nature and its phenomena.
Imagine an experimental physicist in a laboratory, watching a specific physical phenomenon unfolding itself in front of his eyes. Imagine now an identical universe, which however is translated by some amount, and rotated by some angle, and dilated by some scale factor with respect to this universe. Would anything different from the first universe be observed by the physicist in the lab watching the phenomenon he was interested in? In other words, can the experimenter tell in which of these two possible universes he finds himself/herself? In fact the physicist is fully blind to all of these global operations. By moving all the objects in the universe 1 meter to the left, the distances between the objects, would not change at all. That's why the physicist would never see (measure) any difference concerning his state or the state of his environment or even the universe, and intuitively one expects no difference in the way the phenomenon would unfold in front of him. From an internal point of view the universe seems exactly the same, whether it is located here or 1 meter to its left. It looks exactly the same after a total rotation of the universe by some degree, or scaling of the universe (hence all the inter particle distances) with some constant. One could object that by scaling the universe the distances between the objects would also get scaled and hence an internal observer would be able to observe this difference. But as length measurements require rulers, and the inter particle distances of the rulers are scaled up by the same factor as all other distances, one doesn't notice any difference. So: two configurations of the universe which can be transformed into each other by a member of the similarity group $Sim(3)$ are kinematically indistinguishable from an internal point of view. There might be a difference for an external observer, but any discussion on what an external observer of the full universe would see is purely academic, at best it has a philosophical meaning but for sure it is irrelevant for any physical descriptions.  \newline
 
To illustrate the core concepts of the subject matter, let us start with an example of a toy universe which consists of only three particles located in absolute space $\mathbb{R}^{3}$. Each particle's position is given by 3 coordinates, hence 9 numbers are needed to specify the configuration of this system. But as we explained in the last paragraph, this is what an \textit{external} observer watching these three particles in absolute space would say. From an \textit{internal} point of view, for example from the point of view of one of the three particles, not more than 2 degrees of freedom can be observed: the two angles of the triangle formed by these three particles. As explained in the last paragraph, this is because the absolute position (of for instance center of mass), orientation, and scale of the system of 3 particles is unobservable from an internal point of view. One needs 3 numbers to specify the system's center of mass, and 3 numbers to specify the orientation of the system (i.e. Euler angles w.r.t. some frame of reference), and one number to specify the scale of the system. In other words, the similarity group $Sim(3)$ is a $7$ dimensional group, hence its action on the system's configuration space would lead to a $7$ dimensional orbit. As the configuration space of a three particle system was $3\times 3=9$ dimensional, $2$ dimensions remain, which are called the shape degrees of freedom. In general, observations are always internal, thus they  always take place in shape space. Observations always give a quantity in terms of a pre-defined unit of that quantity. Hence the numbers we register as the result of measurements are always relative data and not absolute. \newline
Since from an internal point of view just $2$ angles are observable, one concludes that for our toy universe there exists just $2$ physical degrees of freedom. In other words, the physical configuration space is $2$ dimensional, in contrast to the absolute configuration space which is $9$ dimensional. \newline

This toy model can be indeed generalized to $N$-particles, where $N$ may be as big as the number of elementary particles in the whole universe. The absolute configuration space $Q$ is in that case a $3N$ dimensional, homogeneous space with $Sim(3)$ as its structure group. The fibers $F$ are the orbits (generically 7 dimensional) generated by the action of $Sim(3)$ on absolute configuration space. The base space $S=\frac{Q}{Sim(3)}$ is then isomorphic to the shape space of the universe.\newline $S$ can be understood as the \emph{equivalence classes} of points on absolute configuration space where two points are being set equivalent if and only if they can be transformed into each other by a similarity transformation, i.e. for $x$,$y$ being two points in $Q$ 
\[x\sim y \emph{ if } \exists g\in Sim(3)\mid x=gy\] 
In the physics literature one frequently uses the terms ``relational configuration space'' or \textit{shape space} instead of ``physical configuration space''. How the objects are located with respect to each other, or, equivalently, which shape they form is the only observable from an internal point of view.
Thus formulating the dynamics in this base manifold $S$ is more basic than any description in absolute space. A curve in $S$ corresponds to a unique evolution of the physical degrees of freedom of the system under investigation. These so called physical degrees of freedom are the only quantities visible (or sensible) to internal observers. Adding the gauge degrees of freedom, any curve in $S$ represents an infinite number of trajectories in absolute space\footnote{Note, that a description on phase space, however, often simplifies the equations of motion}, all of which  describe the same phenomena (see discussion above). \newline

Finally, we want to mention that by considering shape space of dimension $3N-7$, we tacitly have assumed the existence of a $3$-dimensional absolute space. In a more general setting one should rather start with a $m$-dimensional shape space where $m$ is not necessarily $3N-7$, and argue how and under which circumstances an apparent 3 dimensional absolute space would emerge for subsystems. Given that the effective 3 dimensionality of absolute space (at least locally) is an empirical fact, we will consider the case $m=3N-7$ in the present manuscript, and postpone  the more general setting to future works.

\subsection{Time as an emergent concept}
According to the worldview of Newton, there exists an absolute 3 dimensional space\footnote{whose existence is independent of matter.}, in which physical objects e.g. particles move. The positions of the particles then change as time passes, and Newton's laws tell how the positions change. Time is an ever flowing external entity, which exists independently of matter and space. In that sense, time generates the dynamics. Without it there is no concept of motion.  
\newline 
Newton did, however, acknowledge that only \textit{relative} positions are experimentally  observable. In the scholium\footnote{A scholium is a explanatory  note in a book written by the author himself.} of his famous book \textit{Principia} \cite{15} he announces to explain how the existence of these absolute structures can be derived from the relative motions of the observable entities. He even claims that this was the central motivation for writing the book \cite{16}, however he does not come back to this later in his book.\newline

Leibniz in contrary was unsatisfied with this way of describing nature.
Accepting that the point like particles of Newtonian Mechanics are the fundamental constituents of the universe, one expects from a physical theory to explain a.o. the behavior (in this case the motions) of these particles. In order to do this, Newton added two extra invisible entities to his description of nature: absolute space and absolute time. Those are essential entities of Newton's laws of motion, specially in his law of inertia. Leibniz in contrast thought of them as extra structures that are unphysical and do not exist in on their own in nature. In his own words \cite{19}:\newline

"\textit{As for my own opinion, I have said more than once, that I hold space to be something merely relative, as time is, that I hold it to be an order of coexistences, as time is an order of successions.}"\newline \newline
Downgrading space to the \textit{order of coexistences}, seems to us a clear denial of Newton's notion of absolute space reviewed above.  
Also  Ernst Mach expressed   in the late 19'th century his critique of absolute time \cite{14}:\newline

"\textit{we must not forget that all things in the world are connected with one another and depend on one another, and that we ourselves and all our thoughts are also a part of nature. It is utterly beyond our power to measure the changes of things by time. Quite the contrary, time is an abstraction, at which we arrive by means of the change of things; made because we are not restricted to any one definite measure, all being interconnected. A motion is termed uniform in which equal increments of space described correspond to equal increments of space described by some motion with which we form a comparison, as the rotation of the earth. A motion may, with respect to another motion, be uniform. But the question whether a motion is in itself uniform, is senseless. With just as little justice, also, may we speak of an absolute time --- of a time independent of change. This absolute time can be measured by comparison with no motion; it has therefore neither a practical nor a scientific value; and no one is justified in saying that he knows aught about it. It is an idle metaphysical conception.}"
\newline\newline
 In a relational theory, time is an emergent concept, and it is most rational to define time in such a way that a change in time always relates to a change in the configuration of the system, for example the change in positions of the atoms forming a pendulum. Time is defined along the trajectories, it is not a concept based on the configurations alone.  Any monotonically increasing function $f$ can be used to define the increment of time via \[\delta t=f(\mid\delta \textbf{x}_{1}\mid,...,\mid\delta \textbf{x}_{N}\mid)\] for infinitesimally small increments of the particle's positions $\mid\delta\textbf{x}_{i}\mid$. Given an arbitrary trajectory on configuration space, including (among others) the two configurations A and B, any monotonous function of the arc length of that trajectory going from configuration A to configuration B can serve, for instance, as a definition of the duration of time passed between A and B. 
As we will review below, duration can be indeed defined as a function of the changes in positions in a unique way such that the Newtonian equations of motion are valid. We are perfectly allowed to use any other suitable function as definition of time, but any different choice  would make the form of the equations of motion different from the ones Newton wrote down, of course; among all possibilities, Newtonian time has the advantage of bringing the equations of motion in its simplest form.
\subsection{Emergence of time in Classical Mechanics}
Originally the principle of least action was developed to justify the equations of motion in different theories. The idea behind the principle was expressed in many different ways by different brilliant thinkers. \textit{Pierre Louis Maupertuis} (1698-1759) is usually credited as the first who gave a concrete formulation of the least action principle (although it is disputed that Leibniz was even earlier). Maupertuis motivation was to rationalize the (by then known) laws of ray optics and mechanics with theological arguments based on design or purpose to explain natural phenomena. Here is a quote from Maupertuis which sheds some light on his views \cite{13}: \newline

"\textit{The laws of movement and of rest deduced from this principle being precisely the same as those observed in nature, we can admire the application of it to all phenomena. The movement of animals, the vegetative growth of plants ... are only its consequences; and the spectacle of the universe becomes so much the grander, so much more beautiful, the worthier of its Author, when one knows that a small number of laws, most wisely established, suffice for all movements}." \newline

Given an initial point in the configuration space of the system under consideration $q_{A}$ and a final point $q_{B}$, the  path chosen by nature minimizes a functional depending on the trajectories connecting the two end points. This functional is usually  called ``action''. In other words: the trajectory  between $q_{A}$ and $q_{B}$ taken  by the system minimizes the value of the action functional. To be more precise, the chosen  trajectory is a stationary point of the action, but in most cases the only stationary point is a minimum.\newline 
Maupertuis proposed to define action as the integral of the so called  \textit{Vis viva} (Latin for 'living force'). The term \textit{Vis viva} was introduced by Leibniz during the 1680's by his observation that the sum of the products of the constituting masses of a system (i.e. a multiparticle system) with the squares of their respective velocities is almost constant during (elastic) collisions, i.e. $\sum_{i=1}^{N}m_{i}v_{i}^{2}=C$. This is, of course,  what we now call principle of conservation of energy (in modern terms Leibniz's \textit{Vis viva} becomes $2$ times the kinetic energy). It seemed to oppose the theory of conservation of momentum advocated by the rival camp (Sir Isaac Newton and Rene Descartes). \newline
Maupertuis' suggestion therefore comes down to the following action functional 
\begin{equation}\label{ma}
W=\int_{t_{A}}^{t_{B}}2Kdt =\int_{q_{A}}^{q_{B}}pdq
\end{equation}  
which is the right formula for systems in which the kinetic energy is quadratic in the velocities. Here the letter $t$ stands for the absolute time of Newton. \newline

\textbf{Maupertuis' principle} states that for the true trajectories (the ones chosen by nature),  Maupertius' action $W$ is stationary on all trial trajectories with fixed initial and final positions $q_{A}$ and $q_{B}$ and fixed energy\footnote{For the most general mechanical systems, energy is $E=\sum_{i}p_{i}\dot{q}_{i}-L$ which reduces to the well known expression $T+V$ in cases where the Lagrangian can be written as $L=T-V$ with $T$ being a quadratic form in the velocities, and $V$ being independent of the velocities.} $E=K+V$.
\begin{equation}\label{mp}
(\delta W)_{E}=0
\end{equation}
where in variational calculus the constraint of fixed end points is usually left implicit, and every other constraint on the trial trajectories (thus in this case fixed energy) is written down explicitly. Note that in \eqref{ma} no constraint is imposed on the value of $t_{B}$, as for different paths a different amount of absolute time is needed to reach the endpoint $q_{B}$. In other words: $t_{b}$ stands for the absolute time (moment) at which the configuration $q_{B}$ is reached, and this obviously changes as the path taken between $q_{A}$ and $q_{B}$ changes. \newline

However, in modern physics literature the most common action principle is the minimization of Hamilton's action denoted by $S$. It is defined as an integral along the \textit{spacetime} trajectory $q(t)$ connecting two configurations $q_{A}=q(t_{A})$ and $q_{B}=q(t_{B})$
\begin{equation}\label{haf}
S=\int_{t_{A}}^{t_{B}}L(q,\dot{q})dt
\end{equation}       
The statement of \textbf{Hamilton's principle} then becomes as following: among all possible trajectories $q(t)$ that can connect the two configurations $q_{A}$ and $q_{B}$ during the exact given time interval $t_{B}-t_{A}=T$, the chosen trajectories are those making $S$ minimal (respectively stationary). 
Thus, Hamilton's principle can be written as
\begin{equation}\label{hp}
(\delta S)_{T}=0
\end{equation}
where, as before, the extra constraint of constant travel time is assumed and denoted explicitly as subscript. Bear in mind that there may be more than one trajectory satisfying these constraints of fixed endpoints and travel time, see \cite{8}. As mentioned, in contrary to  Maupertuis' principle, the allowable trial trajectories of Hamilton's principle do not need to satisfy the constant energy constraint \textit{a priori}: the conservation of energy is here a consequence of Hamilton's principle for time invariant systems (i.e. the Lagrangian does not have an explicit time dependence). Thus, Hamilton's principle \eqref{hp} is applicable to both conservative (time-invariant) and non-conservative systems (i.e. systems that have an explicitly time-dependent Lagrangian due to, for instance, time-dependent potentials $V(q,t)$), while Maupertuis' principle \eqref{mp} is restricted to conservative systems. For conservative systems, one can show that Hamilton and Maupertuis principles are equivalent and related to each other by the famous Legendre transformations. The  results one gets from the action principles are curves that stand for the trajectory of the system. It provides us a manifestly covariant way of describing its evolution. For non-holonomic systems, however, non of these action principles are applicable.\newline

So far it seems that Hamilton's action principle is a more general and powerful principle than Maupertuis. \textit{Carl Gustav Jacob Jacobi} (1804 - 1851) thought so, too. But he wanted to take it one step further by  taking  Newton’s intuition of the existence of an absolute time more seriously by treating time as a variable in the variational calculations. In Newton's spirit, the value of time is as important as the value of the $x$-component of a particle's position for example, or any generalized coordinate. Both of them are absolute, and have physical reality. So, if one wants to apply Hamilton's principle properly, one should not use the absolute Newtonian time $t$ as an independent variable, but in contrary all $n+1$ variables $q_{1}$,...,$q_{n}$,$t$ should be considered as functions of some arbitrary independent variable $\tau$. This enables one to include the variation of $t$ in the variational principle. \newline

Thus we aim to write Hamilton's action principle for a system containing $n+1$ degrees of freedom (see\cite{12}). For consistency from now on in this chapter we denote the derivative with respect to the Newtonian(absolute) time $\frac{\partial}{\partial t}$ by dot, and with respect to the independent variable (used to parametrize the $n+1$ physical degrees of freedom) $\frac{\partial}{\partial \tau}$ by prime. 
Starting with Hamilton's action functional \eqref{haf} for the well known Lagrangian $L(q,\dot{q})$ of classical mechanics, which is the difference of kinetic and potential energy of the system,
and rewriting it in terms of the independent variable $\tau$ we get 
\begin{equation}\label{his}
S=\int_{\tau_{A}}^{\tau_{B}}L(q,\frac{q'}{t'})t'd\tau
\end{equation}         
from which the new Lagrangian (for this new system which has $n+1$ degrees of freedom) can be read off, namely \[L_{new}=Lt'.\] 
Although we arrived at this by a simple mathematical step (change of integration variable in \eqref{haf}), be aware of the huge physical difference between \eqref{haf} and \eqref{his}. In the latter we are varying the space-time curves connecting space-time events $A=(q_{A},t_{A})$ and $B=(q_{B},t_{B})$. \newline
Now as no $t$ appears in $L_{new}$, $t$ is then by definition a cyclic variable. Hence its conjugate momentum 
\[p_{t}=\frac{\partial L_{new}}{\partial t'}= \frac{\partial (Lt')}{\partial t'}=L+(\sum_{i=1}^{n}\frac{\partial L}{\partial \dot{q}_{i}}\frac{\partial \dot{q}_{i} }{\partial t'})t' \]\[= L-(\sum_{i=1}^{n}\frac{\partial L}{\partial \dot{q}_{i}}\frac{q_{i}' }{t'^{2}})t'=L-\sum_{i=1}^{n}p_{i}\dot{q}_{i}\]
is a constant of motion. In the third equation chain rule is used. In the fourth equation we used  $\dot{q}=\frac{q'}{t'}$, hence $\frac{\partial \dot{q}_{i} }{\partial t'}=-\frac{q_{i}' }{t'^{2}}$ . \newline
But the expression derived for $p_{t}$ coincides (up to a minus sign) with the first integral of the Lagrangian equations of motion (for scleronomic systems where the equations of constraints do not have explicit time dependence, see \cite{11}), which is defined in the literature as the total energy $E$ of the system.\newline
So in short, if $t$ is a cyclic variable, (which is the case when the corresponding Lagrangian $L$ of the system we started with  is conservative i.e. $L$ has no explicit time dependence) then 
\begin{equation}\label{ET}
p_{t}=-E
\end{equation}
is a constant of motion. This may as well be considered as an alternative derivation of the theorem of conservation of energy for conservative systems. \newline

It is well known that $n_{c}$ cyclic variables can be eliminated form the variational problem resulting in the reduction of the original variational problem by $n_{c}$ degrees of freedom using the general reduction procedure (see e.g. \cite{12}). For this reason cyclic variables are also called \textit{ignorable variables} in Hamilton's formulation of mechanics. In the present case, the ignorable variable is $t$, and we are interested in  reduction with respect to the variable $t$. The modified Lagrangian becomes

\[\bar{L}_{new}=L_{new}-p_{t}t'=Lt'-p_{t}t'=(L-p_{t})t'\]\[=\sum_{i=1}^{n}p_{i}\dot{q}_{i}t'\]
hence the modified action functional becomes
\begin{equation}\label{moda}
\bar{S}=\int_{\tau_{A}}^{\tau_{B}}\bar{L}_{new}= \int_{\tau_{A}}^{\tau_{B}}\sum_{i=1}^{n}p_{i}\dot{q}_{i}t'd\tau=\int_{\tau_{A}}^{\tau_{B}} 2Kt'd\tau
\end{equation}
where expression \eqref{sub} for the kinetic energy $K$ is used.\newline
 
Note here that because $t'd\tau=dt$, the modified action \eqref{moda} can simply be rewritten as Maupertuis' action \eqref{ma} . But Jacobi's dissatisfaction with Maupertuis' principle was of the same fundamental kind as his dissatisfaction with Hamilton's principle --  with which by the way he started his considerations in the first place. In Maupertuis' action the absolute time $t$ is   used as an independent variable for integration. However in the Newtonian worldview, $t$ itself must be subject of the variational calculation, just like any of the $q_{i}$. It matters at which absolute time $t\in[t_{A},t_{B}]$ a given configuration $q_{C}$ (which locates on the true trajectory somewhere between $q_{A}$ and $q_{B}$) is reached; as much as it matters at which value of the generalized coordinate $q_{i}$ a specific value of some $q_{j}$ is reached ($i\neq j$). It is indeed only to this end that one uses the variational principle at all. So Jacobi's concerns are quite justified for a convinced follower of Newtonian philosophy. \newline

It is worth to emphasize that Jacobi's point (of putting time variable and position variables of a mechanical system on equal footing), is a novel formal difference to the works of his predecessors, and truly finds its fundamental motivation in the Newtonian worldview. However his point does not make a practical difference to Maupertuis' principle, if the system subject of variational calculations is conservative, and the initial value of the absolute time is additionally provided \footnote{which is indeed the case in Jacobi's principle, as you choose a spacetime point as your lower boundary of integration \ref{moda}.}. This is because for conservative systems the motion in time is constraint by \eqref{ET}. So the velocity by which the configuration point of the system moves on its trajectory can be calculated  throughout the whole trajectory, and hence we can calculate exactly at which absolute time any intermediate configuration $q_{C}$ has been reached, and in this way no need for variation of $t$ is remained. A simple calculation can elaborate this point, and leads to a formula for the exact value of the absolute time at an intermediate configuration, i.e. 

\[ t_{C}=t_{A}+ \int_{\tau_{A}}^{\tau_{C}} t'd\tau=t_{A}+\int_{q_{A}}^{q_{C}} \frac{dl}{\sqrt{2K}}\]\[=t_{A}+\int_{q_{A}}^{q_{C}} \frac{dl}{\sqrt{2(E-V)}}\]\[=t_{A}+\int_{\tau_{A}}^{\tau_{C}}\frac{\sqrt{<q'\mid q'>}}{\sqrt{2(E-V)}}d\tau \]
where $q(\tau)$ stands of course for the true trajectory of the conservative system in configuration space (in this case also a solution of Maupertuis' principle), $V=V\big(q(\tau)\big)$, and the norm $<.\mid.>$ on configuration space is defined with respect to the mass tensor (see appendix ii). In the second equality the expression of the kinetic energy in Newtonian theory (i.e. \eqref{cpoint}) is used to substitute the increment of absolute time $dt$ with the line element of configuration space $dl$.\newline
So the use of absolute time as an independent variable in Maupertuis' principle is \textit{a posteriori} satisfied. From Jacobi's analysis it becomes clear that the circumstances under which Maupertuis' principle is applied (constancy of the total energy) leaves no room for a variation of absolute time (its value being fixed for any intermediate configuration as we have just calculated).  \newline

Now let's move on with the last step of the reduction of a cyclic variable, which is the elimination of its velocity using the equation of motion of its conjugate momentum, in this case eliminating $t'$ using \eqref{ET}.\newline 

We equip  configuration space with a Riemannian metric, and set this metric equal to the mass tensor $\textbf{M}$. Then the kinetic energy can simply  be expressed as,
\begin{equation}\label{cpoint}
K=\frac{1}{2}\big(\frac{dl}{dt}\big)^{2}
\end{equation}
 in which $dl$ denotes the line element (with respect to $\textbf{M}$ as metric)\footnote{$dl^{2}=\textbf{M}_{ij}dx^{i}dx^{j}$} of configuration space.  So, in Newtonian theory, the velocity  with which the configuration point of a system moves in configuration space  is $\sqrt{2K}$. Again, in the present case, since we use $\tau$ as our independent variable, \eqref{cpoint} needs to be rewritten as  
\begin{equation}\label{ho}
K= \frac{1}{2}\big(\frac{dl}{d\tau}\big)^{2}/t'^{2}.
\end{equation}    
Now, using this and the momentum equation (which is equivalent to the energy theorem\eqref{ET}),  the cyclic variable (i.e. the remaining $t'$ in the modified action \eqref{moda}) can eventually be eliminated by inserting  
\begin{equation}\label{time}
t'=\frac{1}{\sqrt{2(E-V)}}\frac{dl}{d\tau}
\end{equation}        
  into \eqref{moda}
\[\bar{S}=\int_{\tau_{A}}^{\tau_{B}} 2Kt'd\tau=\int_{\tau_{A}}^{\tau_{B}} 2K\frac{1}{\sqrt{2(E-V)}}\frac{dl}{d\tau}d\tau\]\[=\int_{\tau_{A}}^{\tau_{B}} 2(E-V)\frac{1}{\sqrt{2(E-V)}}\frac{dl}{d\tau}d\tau\]\[=\int_{\tau_{A}}^{\tau_{B}}\sqrt{2(E-V)}\frac{dl}{d\tau}d\tau\]\[=\int_{A}^{B}\sqrt{2(E-V)}dl\]
This finally leads to the reduced action functional 
\begin{equation}\label{Ja}
\bar{S}=\int_{\tau_{A}}^{\tau_{B}}\sqrt{2(E-V)}\frac{dl}{d\tau}d\tau=\int_{A}^{B}\sqrt{2(E-V)}dl
\end{equation}
where the last equation shows invariance of this expression with respect to re-parametrizations. As usual, the minimizing path's of the action satisfy
\begin{equation}\label{jp}
(\delta \bar{S})=0
\end{equation}
and we have arrived at what is called \textbf{Jacobi's principle}. Please note that constancy of the total energy $E$ here is not a constraint imposed manually in the variational calculation (as in \eqref{mp}), but a consequence of Hamilton's principle for time invariant systems; hence we didn't write a letter $E$ explicitly in \eqref{jp}. As is evident from \eqref{Ja}, absolute time $t$ does not appear in its formulation. The solution of this principle is a path in configuration space, without any reference to the motion in absolute time. However, the motion in absolute time (which was of course the question of Jacobi in first place) can easily be recovered from \eqref{time}, the integration of which gives us the physical time $t$ as a function of the independent parameter $\tau$ (which now parametrizes all $n+1$ degrees of freedom as Jacobi wanted).\newline 

Much more recently \cite{1}, Julian Barbour  preferred to rewrite Jacobi's action \eqref{Ja} as 
\begin{equation}\label{jf}
\bar{S}=2\int_{\tau_{A}}^{\tau_{B}}\sqrt{E-V}\sqrt{\tilde{K}}d\tau
\end{equation}  
 where $\tilde{K}:=\frac{1}{2}\frac{dq}{d\tau}.\frac{dq}{d\tau}$ with the inner product defined with respect to the mass tensor, again. It can easily be rewritten as $\tilde{K}=\frac{1}{2}\sum_{i=1}^{N}\frac{d\textbf{r}_{i}}{d\tau}.\frac{d\textbf{r}_{i}}{d\tau}$ where dot denotes the standard Euclidean metric on $\mathbb{R}^{3}$, and the index $i$ runs over the number of particles. Note that $\tilde{K}$ has nothing to do with the kinetic energy which is a Newtonian term. When one uses the physical time to express the velocities, $\tilde{K}$ becomes by definition the kinetic energy we are all familiar with. \newline
The Lagrangian $L$ read off from \eqref{jf} is used to define the canonical momenta 
\begin{equation}\label{momen}
p_{i}:=\frac{\partial L}{\partial (\frac{dq_{i}}{d\tau})}=m_{i}\sqrt{\frac{E-V}{\tilde{K}}}\frac{dq_{i}}{d\tau}
\end{equation}
and the corresponding Euler-Lagrange equation becomes
\begin{equation}\label{babo}
\frac{dp^{i}}{d\tau}=\frac{\partial L}{\partial q_{i}}=-\sqrt{\frac{\tilde{K}}{E-V}}\frac{\partial V}{\partial q_{i}}.
\end{equation}
Remember that we had the full freedom to choose any independent variable $\tau$  since the action \eqref{jf} is reparametrization invariant. One possibility is to choose a parametrization which is such that $\tilde{K}=E-V$. We already know that this specific option for $\tau$ mimics the absolute time of Newton for two reasons. First, inside Newtonian Mechanics for a conservative system, the kinetic energy is of course $E-V$, and as $\tau$ here has been chosen such that $\tilde{K}$ becomes equal to $E-V$, we can conclude that this specific choice for $\tau$ marches in synchronous to absolute time. Another reason which is even more convincing, is that for this specific $\tau$, equation \eqref{babo} takes its familiar form, namely Newton's second law. \newline
Now from $\tilde{K}=\frac{1}{2}\sum_{i=1}^{N}\frac{d\textbf{r}_{i}}{d\tau}.\frac{d\textbf{r}_{i}}{d\tau}=E-V$ one easily deduces 
\begin{equation}\label{ntime}
 dt=\frac{\sqrt{\sum_{i=1}^{N} m_{i}d \textbf{r}_{i}.d \textbf{r}_{i}}}{\sqrt{2(E-V)}}=\frac{dl}{\sqrt{2(E-V)}},
\end{equation}      
where $ds$ is again the line element of the configuration space with respect to the mass tensor. \newline
As Barbour said, from this one can see how \textit{change creates time}. See also \cite{18} for a beautiful presentation of the notion of relational time, from physical and historical point of view.    \newline
The sentence of Mach we quoted at the beginning of this chapter "\textit{It is utterly beyond our power to measure the changes of things by time. Quite the contrary, time is an abstraction, at which we arrive by means of the change of things}", is fully reflected in \eqref{ntime}. \newline
\eqref{ntime} illustrates another fully holistic feature of Newtonian theory which was invisible to us in the way Newton was advocating it. A change in the position of the most far objects, causes the time we are experiencing to move forward! This idea is however not far from how in practice astronomers defined and used the so called \textit{ephemeris time}. The motion of planet earth with respect to the farthest stars was used since thousands of years to  keep track of the time passing, which is a similar concept. Time is a derived notion and not a primitive one as Leibniz was emphasizing all the time. \newline
 Using Jacobi's principle and finally arriving at \eqref{ntime} might have shaken Jacobi in his grave.

\subsection{Principle of Relationalism}
In the first part of this section we have already mentioned the difference between the two alternative worldviews of Newtonian absolutism and Leibnizian relationalism.  Given, that:\newline\newline
1. Most of our current physical theories (like Classical Mechanics, Quantum Mechanics , ...) are based on the Newtonian worldview;\newline
2. Predictions of our current physical theories are compatible with the empirical data to an astonishingly high degree of accuracy, and give a quite clear explanation for the occurrence of numerous natural phenomena;\newline
3. We think relational worldview should be adopted in physics,\newline\newline
 it is  at a  first sight unclear whether a realistic relational theory can be formulated at all, because it is not clear whether statements 2 and 3 are compatible with each other. In the following we will explain how both statements 2 and 3 can be true. To this end we introduce the \textbf{Principle of Relationalism} as follows:
\begin{center}
    Two possible absolute configurations of the universe, differing from each other just by a global similarity transformation $sim(3)$, are observationally\footnote{with observationally indistinguishable we mean kinematically and dynamically indistinguishable.} indistinguishable. 
\end{center}
If a theory based on the Newtonian worldview, satisfies the Principle of Relationalism, it can be recasts into an empirically equivalent theory based on the Leibnizian worldview. \newline

\subsubsection{Scale invariant Classical Gravity}
A natural question to ask now is whether Classical Mechanics satisfies the Principle of Relationalism or not? As the interaction potential functions like Newton's gravitational potential $V=\frac{Gm_{1}m_{2}}{r_{12}}$, or Coulomb potential $V=\frac{1}{4\pi\epsilon_{0}}\frac{q_{1}q_{2}}{r_{12}}$ defined on the absolute space, though being manifestly rotational and translational invariant, are clearly not scale invariant, the answer of the above question seems to be \textit{negative}, and with it the hope for a relational understanding of Classical Mechanics is vanished. But prior to the above question, we should have asked another more primitive question. \textit{Do we already know everything about Newtonian theory of classical Mechanics?} The answer to this question may be ``Yes'', if we had a derivation of the value of for instance the gravitational coupling constant $G$ from Newtonian theory. In other words, we have the opinion that in a complete physical theory based on the Newtonian worldview, there exists a theoretical derivation of the value of the gravitational constant $G$, which then must of course coincide with the value observed in our universe. So even if the Newtonian worldview is the correct view, Newtonian theory of gravitation may very well contain a foundational incompleteness (or gap) in it, as will be explained in more details below in section iv.2 \footnote{specifically a quotation there from Albert Einstein would be illuminating in this matter.}. In the following we propose a way to partially fill this gap in a manner that is compatible with the Principle of Relationalism.\newline

We can always render an arbitrary potential function defined on $Q_{cm}$ or absolute space, scale-invariant, by postulating a special kind of scale-dependent transformation law for its coupling constant. This law should be exactly the inverse of the transformation law of the potential function without its coupling constant. In this way, scale invariance of the total potential function we started with on the absolute space is established.\newline 
So take any potential function \[V(r_{1},...,r_{N})=Yf(r_{1},...,r_{N})\] with $Y$ being its coupling constant. Now apply a scale transformation \[r_{i}\rightarrow c r_{i}\] with $c \in\mathbb{R}^{+}$. Under this transformation the function $f$ and its coupling constant $Y$ will transform as the following
\[\left\{
  \begin{array}{lr}
    f(r_{1},...,r_{N})\rightarrow f'(r_{1},...,r_{N}):=f(c r_{1},...,c r_{N})\\
    Y\rightarrow Y'.
  \end{array}
\right.
\]                 
Then clearly the potential $V$ transforms as \[V\rightarrow V'=Y'f'\] 
Now by requiring $V$ to be scale invariant i.e. $V'=V$, we can deduce the required transformation law of $Y$, namely \[Y'=Y\frac{f}{f'}.\]  
In other words, if $f$ is a homogeneous function of degree $k$ on absolute configuration space (as it is the case for gravitational potential), $Y$ must also be a homogeneous function but of degree $-k$. 
\newline
This way the potential $V(x)$ on absolute space uniquely projects down to a potential function $V_{s}(q)$ on the reduced configuration space $\frac{\mathbb{R}^{3N}}{\mathbb{R}}\cong \mathbb{R}^{3N-1}$ w.r.t. the scale transformations. Denote this projection by $\pi : \mathbb{R}^{3N}\rightarrow \mathbb{R}^{3N-1}$. Then for each $q\in\frac{\mathbb{R}^{3N}}{\mathbb{R}}\cong \mathbb{R}^{3N-1}$ and $x\in \pi^{-1}(q)$ we have 
\begin{equation}\label{spo}
V_{s}(q)=V(x)
\end{equation}
This assignment is indeed independent of $x$ (as long as it lies on that fiber above $q$), because $V(x)$ is a scale invariant function on absolute configuration space.  In other words, to each equivalence class of configurations under scale transformations \footnote{or obviously similarity transformation}, one unique value of the potential is given. Of course to find out the unique value for a given shape, one has to choose a representative of the equivalence class, and this representative may as well be our good old representation in which $G=6.67408\times 10^{-11}m^{3}kg^{-1}s^{-2}$. In this way, we can make classical gravity scale invariant, and hence compatible with the Principle of Relationalism. Equivalently one can say in the gauge where the length of the international prototype meter bar is set to $1$ meter, the measured value of $G$ in our universe\footnote{strictly speaking near Earth} in its current state becomes the above value.   

\subsubsection{Constants of Nature}

 In the last subsection we have introduced a transformation law for the value of the gravitational constant $G$. This is obviously in conflict with the general belief that $G$ is a constant. Hence we found it necessary to clarify this point, and clear up possible confusions which may arise in this regard.  To be more precise, we argue which constants of nature (numbers appearing in laws of nature of the respective theories) have to remain unchanged, given that the universe has to look and work the same way after a global scale transformation. We will see that $G$ is not among those unchanging  "\emph{constants of nature}", neither is the Planck's constant $\hbar$, nor the vacuum permitivity $\epsilon_{0}$.\newline
     
Observationally, we experience a huge number of regularities and fascinating patterns in nature, and our quest to understand the reason of their occurrence, leads us to the discovery of the laws of Nature in which a collection of dimensionless numbers appear whose exact values are not derived in any way inside the theory, but rather determined experimentally. In words of \textit{John D. Barrow} \cite{3} these dimensionless numbers capture at the same time our greatest knowledge and our greatest ignorance about the universe.\newline To make it clear to the reader which numbers we are referring to, the nice correspondence of Einstein with Ilse Rosenthal-Schneider \cite{4} on this topic is very helpful, i.e.\newline

"\emph{Now let there be a complete theory of physics in whose fundamental equations the "universal" constants $c_{1}$,$c_{2}$,...,$c_{n}$ occur. The quantities may somehow be reduced $g$,$cm$,$sec$. The choice of these three units are obviously quite conventional. Each of these $c_{1}$,...,$c_{n}$ has a dimension in these units. We now will choose conditions in such a way that $c_{1}$,$c_{2}$,$c_{3}$ have such dimensions that it is not possible to construct from them a dimensionless product $c_{1}^{\alpha}c_{2}^{\beta}c_{3}^{\gamma}$. Then one can multiply $c_{4}$, $c_{5}$, etc., in such a way by factors built from powers of $c_{1}$,$c_{2}$,$c_{3}$ that these new symbols $c_{4}^{*}$, $c_{5}^{*}$, $c_{6}^{*}$ are pure numbers. These are the genuine universal constants of the theoretical system which have nothing to do with conventional units. My expectation now is that these constants $c_{4}^{*}$ etc., must be basic numbers whose values are established through the logical foundation of the whole theory. Or could put it like this: In a reasonable theory there are no dimensionless numbers whose values are only empirically determinable. Dimensionless constants in the laws of nature, which from the purely logical point of view can just as well have different values, should not exist. To me, with my "trust in god" this appears to be evident, but there will be few who are of the same opinion.}" \newline

To Max Planck it seemed natural that the three dimensionful constants $G$, $\hbar$, $\pmb{c}$ which appear in physical theories  determine the three basic measuring units. The units derived from them retain their natural significance as long as the law of gravitation, and that of propagation of light in a vacuum remain valid. They therefore must be found always to be the same, when measured by the most widely differing intelligences according to the most widely differing methods. He defined the Plank mass and length and time units as 
\[L_{P}=\sqrt{\frac{G\hbar}{\pmb{c}^{3}}}=1.616\times10^{-35}m\]
\[M_{P}=\sqrt{\frac{\hbar \pmb{c}}{G}}=2.177\times 10^{-5}g\]
\[T_{p}=\sqrt{\frac{G\hbar}{\pmb{c}^{5}}}=5.390\times 10^{-44}s\]  
These are Einstein's dimensionful  constants $c_{1}$, $c_{2}$, $c_{3}$. 

So far we are aware of $4$ distinct forces of nature, i.e. gravity, electromagnetism, weak and strong forces. The strength of these 3 forces (compared to the strong force) can be considered  dimensionless (or Einstein's pure) numbers that define our world. The value of these dimensionless numbers are
\[\alpha_{EM}:=\frac{e^{2}}{4\pi\epsilon_{0}\hbar \pmb{c}}\approx \frac{1}{137.036}\]
\[\alpha_{G}:=\frac{Gm_{p}^{2}}{\hbar \pmb{c}}\approx 5\times 10^{-39}\]
\[\alpha_{W}:=\frac{G_{F}m_{p}^{2}\pmb{c}}{\hbar^{3}\sim1.03\times 10^{-5}}\approx 10^{-15}\] 
Universes for which the value of any of these three dimensionless numbers are different from the above values are  observationally different(cause the balances of the forces differ). \newline

Coming back to the proposed transformation law $G\rightarrow c G$ of the gravitational constant under scale transformations $r\rightarrow c r$, one can immediately see from $\alpha_{G}$, that it changes the balance of Forces in Nature, and leads to observable differences, which clearly violates the Principle of Relationalism. However, if this transformation of $G$ is accompanied by a transformation $\hbar\rightarrow c\hbar$ of the Planck's constant, the strength of gravity would remain unchanged. It is well known from Quantum mechanics that even the smallest change of the value of $\hbar$ would lead to a sudden release or absorption of an enormous amount of energy, due to the  dependence of the atomic orbital energy levels to the value of $\hbar$, i.e. for the hydrogen atom $E_{n}=-\frac{me^{4}}{8h^{2}\epsilon_{0}^{2}}\frac{1}{n^{2}}$, and this again violates the Principle of Relationalism. However, if a transformation $\epsilon_{0}\rightarrow \frac{\epsilon_{0}}{c}$ of the vacuum permitivity is also taking place along the mentioned transformations of $G$ and $\hbar$, the value of energy levels remain unchanged so that the mentioned principle is respected.\newline

To summarize, after performing a scale transformation by a factor $c\in\mathbb{R}^{+}$ 
\[r\rightarrow c r\] on the whole universe, the Principle of Relationalism requires the following transformation of (Einstein's dimensionful) constants
\begin{equation}
G\rightarrow c G
\end{equation}  
\begin{equation}\label{habs}
\hbar\rightarrow c\hbar
\end{equation} 
\begin{equation}
\epsilon_{0} \rightarrow  \frac{\epsilon_{0}}{c}
\end{equation} 
To appreciate the consistency of these transformation laws more, notice that they automatically induce the expected transformation of the Bohr radius $a_{0}=\frac{4\pi\epsilon_{0}\hbar^{2}}{m_{e}e^{2}}$, classical electron's radius $r_{e}=\frac{1}{4\pi\epsilon_{0}}\frac{e^{2}}{m_{e}\pmb{c}^{2}}$, and the fine structure constant $\alpha_{EM}$, namely $a_{0}\rightarrow ca_{0}$, $r_{e}\rightarrow cr_{e}$ and $\alpha_{EM}\rightarrow \alpha_{EM}$. So all atoms in the universe get bigger by exactly the same scaling factor for the universe itself and the strength of the electromagnetic force remains unchanged. Moreover they also automatically result in a scale invariant electrical force (Coulomb's potential). \newline

Imagine an experiment by which you want to figure out the velocity of an object. We argue that after performing of a global scale transformation, the velocity of the same object during the same experiment remains unchanged. One can see this in the following way: From the formulas of Planck's system of units mentioned above, one can easily see that a scale transformation \[r\rightarrow c r\] reshuffles the values of these units expressed in SI\footnote{Here SI is being thought of as some measures of hypothetical absolute space and time, in the sense that they are exempted from the global transformation we perform on the universe. So we scale everything in the universe except the SI standard meter stick and the SI standard clock as resembling the absolute distance and absolute time duration, and a hypothetical mean to compare the new Plank's units after expansion to the old ones inside the absolute framework that classical theories and quantum mechanics are presented.} as 
\[L_{p}\rightarrow c L_{p}\] \[M_{p}\rightarrow M_{p}\]  \[T_{p}\rightarrow c T_{p}.\] 
As expected the natural unit of length gets bigger by the same factor  $c\in \mathbb{R}^{+}$. The time (measured in Planck's unit) gets also dilated and runs faster by the same factor.\newline 
Hence the measured speed $v$ of an object transforms under a global scale transformation as follows 
\[v=\frac{\Delta x}{\Delta t}\rightarrow v'=\frac{\Delta x'}{\Delta t'}=\frac{c\Delta x}{c\Delta t}=v\] 
where $\Delta x$ stands for instance for the distance between two other objects (which are needed to define the start and end point of any interval in space), and $\Delta t$ for the time (measured in Plank unit) the object takes to travel between those two reference objects. The primed versions have the same quantities, however after scale transforming the universe and measuring everything in new Plank units. The same can be said about the velocity of light \footnote{To be more precise, the average two way light's velocity is meant here. There exist no experimental way to measure the direct one way velocity of light as a consequence of conventionality of simultaneity.}  \textbf{c}, where one measures the time needed for light to path the distance between two objects. Note that in the relation $\Delta t'=c \Delta t$ the Principle of Relationalism is tacitly invoked in equating the number of ticks(or steps) of our new clock in the scaled universe for the duration of a physical phenomenon (in this example the passage of light of an object between the two reference objects), and the number of ticks of the old clock in the old (smaller) universe while the same phenomenon is taking place. So the measured speed of any object in universes before and after global scale transformations comes out the same. The Principle of Relationalism, and the characteristics of Plank units together are responsible for this result.\newline 
As a consequence of the constancy of the speed of light (measured in Planck units) $\textbf{c}=\frac{1}{\sqrt{\epsilon_{0}\mu_{0}}}$ under scale transformations, one can deduce the corresponding transformation law of vacuum Permeability, namely
\[\mu_{0}\rightarrow c\mu_{0}\]  
The dilation of time under scale transformation in the presented way is also compatible with operational SI definition of time unit, i.e. the \emph{second} is defined as the duration of $9192631770$ cycles of radiation corresponding to the transition between two energy levels of the ground state of the cesium-133 atom at rest at a temperature of absolute zero. By performing a scale transformation $r\rightarrow c r$, the wave-length of the emitted photon transforms correspondingly ($\lambda_{photon}\rightarrow c\lambda_{photon}$), and hence the time required for one cycle , i.e.  $T=\frac{\lambda_{photon}}{\textbf{c}}$, transforms as $T\rightarrow c T$. In this way the SI second will also get dilated by the same factor $c$. This shows the expected coherence between Planck and SI units of time under global scalings. \newline

Remember that Jacobi's principle stated that the path taken by a classical system minimizes the Jacobi action $\bar{S}=\int_{x_{1}}^{x_{2}}\sqrt{E-V}ds$ with $x_{1}$ and $x_{2}$ standing for the initial and final configuration of the system. For the path $\textbf{x}(t)$ that minimizes this action one has the energy conservation equation i.e. $E=\frac{1}{2}M(\frac{d\textbf{x}}{dt},\frac{d\textbf{x}}{dt})+V$. Hence along this path (which is the only physical path in the sense that only this path is realized by nature) one has $K:=\frac{1}{2}M(\frac{d\textbf{x}}{dt},\frac{d\textbf{x}}{dt})=E-V$. Now Jacobi action along this path can be rewritten as $\bar{S}=\int_{x_{1}}^{x_{2}}\sqrt{K}ds$. If one now performs a scale transformation $r\rightarrow c r$, with $c\in \mathbb{R}^{+} $ the system naturally gets bigger but the velocity of the constituting particles of the system measured in the new Planck units of time and length remain unchanged. Moreover, the length of the path between $cx_{1}$ and $cx_{2}$ measured in the new Planck length remains also unchanged. So in this way one sees that the action of classical mechanics is invariant under scale transformations.

\newpage
\section{Symplectic reduction of phase space with respect to a symmetry group}
In this section we will review how the collective motion of a multi-particle system (in
particular the rotations and translations of system) is gotten rid of in the Hamiltonain formalism using the reduction procedure of Mardsen/Weinstein. We follow \cite{5},\cite{6} to a big extend. First we review the expression of the laws of classical mechanics using a symplectic structure on phase space.
This level of abstraction for formulating Classical Mechanics seems at first sight to be an unnecessary complication, but its power lies in its generality, and is beneficial, compared to less abstract formulations, when dealing with curved spaces. This is the case in many  mechanical systems with constraints, as well as on reduced spaces like internal configuration space $Q_{int}=\frac{Q}{E(3)}=\frac{\mathbb{R}^{3N}}{E(3)}$ or shape space $S=\frac{Q}{sim(3)}$ of a $N$-particle system. 

\subsection{Definition of a Hamiltonian system in symplectic phase space}
Denote the configuration space of a $N$-particle system with $Q\cong \mathbb{R}^{3N}$. A symplectic form $\sigma$ on $Q$ is a closed non-degenerate differential two-form. Closed means that the exterior derivative of $\sigma$ vanishes, i.e. $d\sigma=0$, and non-degenarate means that if there exists some $u\in T_{x}(Q)$ such that $\sigma(u,v)=0$ for all $v\in T_{x}(Q)$, then $u=0$. The Hamiltonian $H$ is a function on $T^{*}(Q)$ to which one can associate the respective Hamiltonian flow, which is a vector field $X_{H}$ on $Q$ defined by the equation $\sigma(X_{H},Y)=dH$ for all $Y\in T(Q)$. Symplectic geometry is well suited for investigating mechanical systems. Starting with the configuration space $Q$ of a system, its phase space $T^{*}(Q)$ is canonically symplectic. Denoting the configuration coordinates by $q_{i}$, and the remaining coordinates needed on $T^{*}(Q)$ by $p_{i}$, the canonical symplectic form becomes $\sigma=\sum_{i=1}^{N}dq_{i}\wedge dp_{i}$. The Hamiltonian flow associated to a physical Hamiltonian $H=\sum_{i=1}^{N}\frac{p_{i}^{2}}{2}+V$ leads to an evolution of the system's initial state $(q_{i}(0),p_{i}(0))\in T^{*}(Q)$, which is compatible with Newton's laws of motion. In the following we explain this construction more precisely.\newline

The cotangent space $T^{*}_{x}(Q)$ at $x\in Q$ is isomorphic to the tangent space $T_{x}(Q)$ by the induced isomorphism defined through the following equation
\begin{equation}\label{pvco}
I_{x}:T_{x}(Q)\rightarrow T^{*}_{x}(Q)
\end{equation}
\[ I_{x}(v).u=K_{x}(u,v) \]
for $u,v\in T_{x}(Q)$. Here $.$ stands for the pairing of vectors $T(Q)$ and covectors $T^{*}(Q)$.\newline
Setting $p:=I_{x}(v)$ and writing $p=(p_{1},p_{2},...,p_{N})$ as a tuple, we get from the definition of $K_{x}$
\begin{subequations}
\begin{align}
        p_{k}=m_{k}v_{k}\\
        p.u=\sum_{k=1}^{N} (p_{k},u_{k}).
\end{align}
\end{subequations}
Thus we have obtained the induced variables $x$ and $p$, constituting a coordinate system of the cotangent space $T^{*}(Q)\cong Q\times \mathbb{R}^{3N}$. $x$ and $p$ are often called the coordinate and momentum variables.\newline
Now we define the canonical one-form $\theta$ on the cotangent bundle $T^{*}(Q)$. For \[(u,w)\in T(T^{*}(Q))\] being a tangent vector at \[(x,p)\in T^{*}(Q)\cong Q\times \mathbb{R}^{dN}\] we define
\begin{equation}
\theta_{(x,p)}(u,w):=p.u\;.
\end{equation}
If $u$ is a vector field on $Q$, then $dx^{i}_{k}(u)=u^{i}_{k}$ in Cartesian coordinates, so that the canonical one-form $\theta$ can be expressed in the following form
\begin{equation}
\theta = p.dx = \sum (p_{k},dx_{k}).
\end{equation}
The exterior derivative of $\theta$ reads \[d\theta=dp\wedge dx=\sum (dp_{k}\wedge dx_{k}).\]
A scalar product $K^{*}_{x}$ on the cotangent space $T^{*}_{x}(Q)$ can be defined as
\begin{equation}
K^{*}_{x}(q,p):=K_{x}(I^{-1}_{x}(q),I^{1}_{x}(p))=\sum\frac{(q_{k}\mid p_{k})}{m_{k}}
\end{equation}
for $q,p\in T^{*}_{x}(Q)$.\newline
The Hamiltonian of a system is a function on $T^{*}(Q)$ of the following form
\begin{equation}
H=K^{*}(p,p)+U,
\end{equation}
where $U$ is a potential function invariant under translations  ($\mathbb{R}^{n}$) and rotations ($SO(3)$).\newline 
The triple \[(T^{*}(Q),d\theta,H)\] constitutes a Hamiltonian system.\newline 
Hamilton's equations of motion are given by the Hamiltonian vector field $\textbf{X}_{H}$, which is defined through \[d\theta(\textbf{X}_{H},\textbf{Y})=dH(\textbf{Y})\] $\forall \textbf{Y}\in T(Q)$.\newline

Before moving to the next section, we shortly discuss how a symplectic form can be used to express electromagnetic laws of motion for a charged particle \cite{9} for the purpose of illustration. Using the isomorphism \eqref{pvco}, the magnetic vector potential $\textbf{A}$ can be considered as a $1$-form. Then the magnetic field $\textbf{B}$ becomes a 2-form $d\textbf{A}$ on configuration space $Q$. Gauss's law for magnetism  $\nabla .\textbf{B}=0$ is in this formalism expressed as $d\textbf{B}=0$. Then one defines a new phase space $(T^{*}(Q),\sigma_{\textbf{B}})$ which differs from the previous phase space $(T^{*}(Q),\sigma)$ in that the canonical symplectic from $\sigma$ on $T^{*}(Q)$ is replaced with \[\sigma_{\textbf{B}}=\sigma+\pi^{*}\textbf{B},\] 
where $\pi:T^{*}(Q)\rightarrow Q$. Denoting the electric potential function by $\phi$, and denoting the velocity or momentum of the charged particle (which are identified to each other by the metric) by $\textbf{v}$, the Hamiltonian becomes $H=\frac{1}{2}\mid\mid\textbf{v}\mid\mid^{2}+\phi$. The Hamiltonian vector field which defines the dynamics in the presence of Electromagnetic field, can then be derived from $\sigma_{\textbf{B}}(\textbf{X}_{H},\textbf{Y})=dH(\textbf{Y})$ for all $\textbf{Y}\in T(Q)$.       

\subsection{Momentum mappings}
The symplectic structure on $T^{*}(Q)$, enables us to express Noether's theorem more naturally. The action of a Lie-group $G$ on $T^{*}(Q)$ can be generated by a vector field $\textbf{a}_{x}$ on $T^{*}(Q)$, known as infinitesimal generator of the action. Integral curves of $\textbf{a}_{x}$ are the $G$-orbits on $T^{*}(Q)$. Noether's theorem then ensures the existence of a function $\mu$ on $T(Q)$, preserved by the action and conjugated to $\textbf{a}_{x}$ by the symplectic form, i.e. $\sigma(\textbf{a}_{x},Y)=d\mu(Y)$ for all vector fields $Y$. This function is called the momentum map, and it is preserved by the Hamiltonian flow.    
Here we review the important concept of momentum mapping which will be used frequently in the process of reduction.\newline
If a group $G$ acts on the manifold $Q$ and $(.,.)$ is a $G$-invariant Riemannian metric, we define for \[\textbf{a}\in \textbf{G}\] \[v_{x}\in T_{x}Q\] the momentum map $\mu$  as follows
\begin{subequations}
\begin{align}
        \mu :T(Q)\equiv T^{*}(Q)\rightarrow \textbf{G}^{*}\\
        \mu (v_{x}).\textbf{a}:=(\textbf{a}_{x},v_{x})\\ 
        \textbf{a}_{x}=\frac{d (e^{t\textbf{a}}x)}{dt}\mid_{t=0} \in T^{*}_{x}(Q).
\end{align}
\end{subequations}
$T(Q)$ and $T^{*}(Q)$ are identified with the metric.\newline
There is an intrinsic formulation of the connection form in terms of the momentum map. Remember the definition of the inertia tensor(or operator) $A$. It was a linear operator in $\wedge^{2}(3)$, and there existed an isomorphism $R$ between $\wedge^{2}(3)$ and $\textbf{so}(3)$ (see appendix iv), and since the tangent space and the cotangent space of $Q$ are identified through the Riemmanian metric on $Q$, we are able to redefine the inertia operator as follows
\begin{subequations}
\begin{align}
       A:\textbf{G}\rightarrow \textbf{G$^{*}$}\\
        A_{x}(\textbf{a}).\textbf{b}=(\textbf{a}_{x},\textbf{b}_{x}).
\end{align}
\end{subequations}
The connection form is then 
\begin{equation}\label{rotconnfor}
\omega(v_{x})=A^{-1}_{x}(\mu(v_{x})).
\end{equation}
The horizontal distribution is the kernel of the momentum map $\mu$. \newline 
Alternatively, one can also think of $G$ as the group of symplectic transformations (preserving $d\theta$), and $\textbf{G}$ as the Lie-algebra of $G$ (which is identified with the tangent space to $G$ at the identity). For every $\textbf{a}\in \textbf{G}$ we get a one-parameter subgroup of $G$ by $exp(t\textbf{a})$. \newline
If for any $\textbf{a}\in \textbf{G}$ there exists a function $F_{\textbf{a}}$ on $T^{*}(Q)$ satisfying $d\theta\textbf{a}_{x}=-dF_{\textbf{a}}$, then the action of \textbf{$G$} is called strongly symplectic. The function $F_{\textbf{a}}$ depending linearly on $\textbf{a}$, can be expressed in the form $F_{\textbf{a}}(x,p)=\mu(x,p).\textbf{a}$ which is the defining property of the momentum map $\mu$ of $T^{*}(Q)$ to \textbf{$G^{*}$}.\newline
If the action of $G$ is moreover exactly symplectic, $G$ leaves $\theta$ invariant -- then there is a simple equation which gives us the momentum map
\begin{equation}
\mu(x,p).\textbf{a}=\theta (\textbf{a}_{x}).
\end{equation} 
As we see below momentum mappings cover  linear and angular momentum.\newline\newline
It is well-known and intuitively clear that the expression of the connection form (\ref{rotconnfor}) for the $SO(3)$ fiber bundle, in Jacobi coordinates $\pmb{r}_{i}$ becomes 
\begin{equation}\label{rotconnfor1}
    \omega=R\bigg(A_{x}^{-1}(\sum_{j=1}^{N-1}\pmb{r}_{j}\times d\pmb{r}_{j})\bigg)
\end{equation}
where $d\pmb{r}_{j}$ is the 3 dimensional vector valued one form, which, if applied to a vector on configuration space $Q$, gives the velocity vector of just the $j's$ particle.

\subsection{Marsden-Weinstein method of Reduction of dynamical systems}
Consider a symplectic manifold $P=T^{*}(Q)$, the symplectic form $\sigma$  on this manifold, and a  $\sigma$-preserving  symplectic group $G$ acting on $P$. The adjoint $Ad_{g}$ and coadjoint $Ad^{*}_{g}$ representations of $G$ on the Lie-algebra space \textbf{G} and its dual space \textbf{G$^{*}$} respectively, are defined in appendix iii (see \eqref{adjact}).\newline
Let $\mu$ be the $Ad^{*}$-equivariant momentum mapping associated with the action of $G$. That is 
\begin{subequations}
\begin{align}
       \mu:P\rightarrow \textbf{G$^{*}$}\\
        \mu(gx)=Ad^{*}_{g^{-1}}\mu(x), \forall x\in P.
\end{align}
\end{subequations}
For $r\in\textbf{G$^{*}$}$, $\mu^{-1}(r)$ is a submanifold of $P$. The isotropy subgroup $G_{r}$ of $G$ at $r\in\textbf{G$^{*}$}$ is defined as the following
\begin{equation}
G_{r}=\{\forall g\in G\mid Ad^{*}_{g^{-1}}r=r\}.
\end{equation}  
Define then the manifold 
\begin{equation}
P_{r}:=\frac{\mu^{-1}(r)}{G_{r}}
\end{equation}
with its canonical projection 
\begin{equation}
\pi_{r}:\mu^{-1}(r)\rightarrow P_{r}.
\end{equation}
$P_{r}$ is called the \textbf{reduced phase space}.  \newline
With the help of the inclusion map 
\[i_{r}:\mu^{-1}(r)\rightarrow P\] 
we can get a unique symplectic form $\sigma_{r}$ on $P_{r}$
\begin{equation}
\pi^{*}_{r}\sigma_{r}=i^{*}_{r}\sigma.
\end{equation}
And at last, if the Hamiltonian $H$ on $P$ is invariant under the action of $G$, the Hamiltonian vector field $\textbf{X}_{H}$ projects to a vector field $\textbf{X}_{H_{r}}$ on $P_{r}$, namely 
\begin{equation}
\pi_{r_{*}}\textbf{X}_{H}=\textbf{X}_{H_{r}}.
\end{equation}
With 
\begin{equation}
\pi^{*}_{r}H_{r}=i^{*}_{r}H.
\end{equation}
Hence one obtains the \textbf{reduced system} \[(P_{r},\sigma_{r}, H_{r}).\]

\section{Example: Reduction with respect to the Euclidean group $E(3)$}
As an illustration of the general symplectic reduction procedure discussed in the previous section, we review the reduction of phase space of a classical system with respect to the Euclidean group, following \cite{5} to a big extend.    
\subsection{Reduction with respect to the translation group $G=\mathbb{R}^{3}$}
The translation group $\mathbb{R}^{3}$ forms an exact symplectic group on $T^{*}(Q)$. Any member of this group $a\in\mathbb{R}^{3}$ acts on $T^{*}(Q)$ as following which obviously leaves the one-form $\theta$ invariant\newline

\[(x_{1},...,x_{N},p_{1},...,p_{N})\]\[\downarrow\] \[(x_{1}+a,...,x_{N}+a,p_{1},...,p_{N}).\]
For $\textbf{a}\in \mathbb{R}^{3}$, where $\mathbb{R}^{3}$ now stands for the Lie-algebra of the translation group, the infinitesimal generator of the subgroup $a(t)=exp(t\textbf{a})$  has the form \[\textbf{a}_{x}(x,p)=(\textbf{a},...\textbf{a},0,...,0)\in T^{*}_{x}(Q)\] so that the momentum map $\mu_{t}: T^{*}(Q)\rightarrow \mathbb{R}^{3}$ is given by:\newline
$\mu_{t}(x,p).\textbf{a}=\theta(\textbf{a}_{x})=\sum(p_{k},\textbf{a})=(\sum p_{k}\mid \textbf{a})\Rightarrow$ 
\begin{equation}
\mu_{t}(x,p)=\sum p_{k}.
\end{equation}
This way we obtain the usual linear momentum. \newline
Now in order to perform the reduction of the phase space $T^{*}(Q)$ with respect to the translation group $\mathbb{R}^{3}$ we apply the Marsden-Weinstein method. For $\lambda \in \mathbb{R}^{3}$, $\mu^{-1}_{t}(\lambda)$ is a submanifold of $T^{*}(Q)$ determined by $\sum p_{k}=\lambda$. This submanifold is isomorphic with $Q\times \mathbb{R}^{(N-1)3}$ for any $\lambda$. It is clear that the isotropy subgroup at $\lambda$, denoted by $G_{\lambda}$, is the whole group of translations $\mathbb{R}^{3}$. So the reduced phase space \[P_{\lambda}=\frac{\mu^{-1}_{t}(\lambda)}{\mathbb{R}^{3}}\] can be identified with $\frac{Q}{\mathbb{R}^{3}}\times \mathbb{R}^{3(N-1)}$, and therefore with  \[P_{\lambda}\cong Q_{cm}\times \mathbb{R}^{3(N-1)}.\] 
This reduced space can in turn be thought of as a submanifold of $T^{*}(Q)$ determined by the following conditions
\begin{subequations}\label{condi}
\begin{align}
       \sum m_{k}x_{k}=0\\
       \sum p_{k}=\lambda.
\end{align}
\end{subequations} 
What we are interested in is the case $\lambda=0$. The submanifold $\frac{\mu^{-1}_{t}(0)}{\mathbb{R}^{3}}$ can then be identified with the cotangent bundle $T^{*}(Q_{cm})$
\[P_{\lambda=0}\cong T^{*}(Q_{cm}).\]
The reduced symplectic form on $T^{*}(Q_{cm})$ is then the restriction of $d\theta$ (which was the form on $T^{*}(Q)$) on $T^{*}(Q_{cm})$. For notational convenience both of them are denoted by the same letter. So one arrives at the reduced Hamiltonian system with respect to the group of 3 dimensional spatial translations i.e. $(T^{*}(Q_{cm}),d\theta,H)$. \newline

Note that the identification between the reduced phase space with respect to the group of translations $P_{\lambda}=\frac{\mu^{-1}_{t}(\lambda)}{\mathbb{R}^{3}}$, and the cotangent bundle of the center of mass system $T^{*}(Q_{cm})$ holds only for $\lambda=0$. In the remaining part of this subsection we explain one way to see this point more clearly.\newline
 Consider a generic point $y\in T^{*}(Q)$. This point can symbolically be denoted as $y=\left(\begin{array}{c}\overrightarrow{p}_{1}\\...\\ \overrightarrow{p}_{N} \end{array}\right)(x)$ where $\overrightarrow{p}_{i}$ stands for the momentum of $i$'th particle (here the isomorphism between 1-forms and vectors is invoked too), and $x\in Q\cong \mathbb{R}^{3N}$ stands for a point in the configuration space of the multiparticle system. One can view $T^{*}(Q)$ simply as a $2N\cdot3$ dimensional space, which is coordinatized by $\overrightarrow{x}_{1},...,\overrightarrow{x}_{N},\overrightarrow{p}_{1},...,\overrightarrow{p}_{N}$. As each of these vectors consists of $3$ numbers, they are indeed a collection of $6N$ numbers. As explained in the last paragraph, $\frac{\mu^{-1}_{t}(0)}{\mathbb{R}^{3}}$ can be considered as a $2(N-1).3$ dimensional submanifold of $T^{*}(Q)$ given by the constraints \eqref{condi}. So far so good. One can alternatively view $T^{*}(Q)$ as follows: take the configuration space $Q$ of the system, and attach to each point $x\in Q$ a $(3N)$-dimensional vectorspace. This vector space is thought to be the collection of all possible elements $\left(\begin{array}{c}\overrightarrow{p}_{1}\\...\\ \overrightarrow{p}_{N} \end{array}\right)$ and denote this vector space by $V_{Q}$. Now comes the tricky point. The condition (\ref{condi}.a) gives us a fixed $3(N-1)$ dimensional surface in the absolute configuration space Q. By definition this solid surface can be identified by $Q_{cm}$ (one can even call this surface $Q_{cm}$ no matter whether it is embedded into some bigger space or not). For the moment we denote this surface by $Q^{p}_{cm}$, where $p$ reminds us that this surface is part of a bigger configuration space $Q$. Now  condition (\ref{condi}.b) selects a subspace of the vectorspace which was attached to each point on $Q$, hence also to each point on the surface $Q^{p}_{cm}$. Clearly this subvectorspace consists of special elements $\left(\begin{array}{c}\overrightarrow{p}_{1}\\...\\ \overrightarrow{p}_{N} \end{array}\right)$; namely the ones with $ \sum \overrightarrow{p}_{k}=\lambda$. We denote this subvectorspace by $V_{\lambda}\subset V_{Q}$. In this way $\frac{\mu^{-1}_{t}(\lambda)}{\mathbb{R}^{3}}$ (which again was a submanifold of $T^{*}(Q)$ realized by constraints \eqref{condi}) can be viewed as $Q^{p}_{cm}\times V(\lambda)$.\newline
On the other hand, the cotangent space over the centre of mass system i.e. $T^{*}(Q_{cm})$ is on its own an independently existing $2(N-1)d$ dimensional space, without any need of an ambient space. Similarly this space $T^{*}(Q_{cm})$ can be viewed as $Q_{cm}\times V_{cm}$, where $V_{cm}$ is just the $3(N-1)$ dimensional vector space attached to each point of the center of mass configuration space $Q_{cm}$.\newline 
Now if one tries to embed $T^{*}(Q_{cm})$ into $T^{*}(Q)$ one can indeed perfectly fit $Q_{cm}$ on $Q^{p}_{cm}$, but one can never fit $V_{cm}$ on $V_{\lambda}$ unless $ \lambda =  0 $. The reason is that $V_{\lambda}\bigcap V^{p}_{cm}=\oslash$. Here $V^{p}_{cm}$ denotes the embedding of $V_{cm}$ in $V_{Q}$. Any element $v\in V_{cm}$ will assign a  set of velocities to the particles. Pulled up to the absolute space in the center of mass system (so condition \eqref{condi}.a being valid), these velocities  add up to zero (so they have to be elements of $V_{\lambda=0}$); otherwise we would immediately move out of the surface $Q^{p}_{cm}$ and that results does not fit with $T^{*}(Q_{cm})$ on $Q^{p}_{cm}\times V_{\lambda}$ for $\lambda \neq 0 $.  

\subsection{Reduction with respect to the Rotation group $G=SO(3)$}
We now proceed to the angular momentum defined on $T^{*}(Q)$. The rotation group $SO(3)$  plays here the role of an exact symplectic group (preserving $d\theta$) whose action on $T^{*}(Q)$ is defined for $(x,p)$ and $g\in SO(3)$ by
\begin{equation}\label{actr}
(x,p)\rightarrow (gx,gp).
\end{equation}
For the case of vanishing linear momentum, i.e. $\lambda=0$, we note that $SO(3)$ acts actually on $T^{*}(Q_{cm})$ as the conditions \eqref{condi} are invariant under $SO(3)$. If $\lambda$ is non-vanishing, only a subgroup of $SO(3)$ acts on $\frac{\mu^{-1}_{t}(\lambda)}{\mathbb{R}^{3}}$.\newline

Consider some $\textbf{a} = R_{\xi}\in\textbf{so}(3)$, where $\xi\in\wedge^{2}(3)$ is the two-vector corresponding to the Lie-algebra element $\textbf{a}$, and the correspondence is given by the isomorphism $R:\wedge^{2}(3)\rightarrow \textbf{so}(d)$ introduced in appendix iii, see \eqref{isoR}. The infinitesimal generator of the subgroup $exp(t\textbf{a})$, is given by
\[\textbf{a}_{x}(x,p)=\big(R_{\xi}(x),R_{\xi}(p)\big)\]\[=\big(R_{\xi}(x_{1}),...,R_{\xi}(x_{N}),R_{\xi}(p_{1}),...,R_{\xi}(p_{N})\big)\]
where \eqref{isoR1} is used.
Therefor the momentum mapping \[\mu_{r}: T^{*}(Q_{cm})\rightarrow \textbf{so}^{*}(3)\] can be calculated as follows:
\[\mu_{r}(x,p).\textbf{a} =\theta_{(x,p)}(\textbf{a}_{x}) = \sum_{k=1}^{N}(p_{k}\mid R_{\xi}(x_{k}))\]\[=(\sum_{k=1}^{N} p_{k}\wedge x_{k}\mid\xi)=(R_{\sum_{k=1}^{N} p_{k}\wedge x_{k}}\mid R_{\xi})\] 
where \eqref{properties of R A}.e has been used, and in the last equality the fact that the mapping $R$ is isometric is being invoked. Hence one ends up with  
\begin{equation}
\mu_{r}(x,p)=R_{-\sum_{k=1}^{N} x_{k}\wedge p_{k}}.
\end{equation} 
Here we have identified $\textbf{so}(d)$ and $\textbf{so}^{*}(d)$ through the scalar product on $\textbf{so}(d)$, namely $(\pmb{\alpha},\pmb{\beta})=\frac{1}{2}tr(\pmb{\alpha\beta}^{T})$.\newline
One can prove that for an exact symplectic group (transformations which leave the 1-form $\theta $ invariant), the associated momentum mapping is $Ad^{*}$ -equivariant.\newline\newline
Now we apply the reduction theorem to the rotation group $SO(3)$.\newline
Let $\textbf{a}\in \textbf{so}(3)\cong \textbf{so}^{*}(d)$. Then $\mu^{-1}_{r}(\textbf{a})$ is a submanifold of $T^{*}(Q_{cm})$. Factoring out the orbits of the isotropy subgroup $G_{\textbf{a}}$ of $SO(3)$ at $\textbf{a}$, we obtain a reduced phase space $\frac{\mu^{-1}_{r}(\textbf{a})}{G_{\textbf{a}}}$. This process is merely an elimination of the angular momentum.\newline\newline
An important question now pops up: is the reduced phase space $\frac{\mu^{-1}_{r}(\textbf{a})}{G_{\textbf{a}}}$ diffeomorphic to the cotangent bundle $T^{*}(Q_{int})$ of the internal space $Q_{int}=\frac{Q}{\mathbb{R}^{3} \circ SO(d)}$ as it was the case for the translations group?\newline 
The answer is NO, for $a \neq 0 $. \newline 
For the N-body problem in $\mathbb{R}^{3}$ the dimension of the phase space reduces by 4 when eliminating the angular momentum. This is because the Lie-algebra $\textbf{so}(3)$ is 3 dimensional, and the isotropy subgroup $G_{\textbf{a}}$ for $a \neq 0 $ turns out to be $SO(2)$. So the condition $\mu_{r}= \textbf{a}$ in phase space diminishes the dimension by 3, and factoring out the $SO(2)$ orbits does by 1. Intuitively, once a single member of the 3-dimensional space $\textbf{so}(3)$ has been chosen  for the total angular momentum $\textbf{a}$ of the Hamiltonian system, we end up on a sub-manifold with 3 dimensions less. Now, in the original space start rotating the whole system about an axis which is parallel to the total angular momentum vector and passes through the center of mass of the system. This is indeed an $SO(2)$ rotation. It is clear that the value of the momentum map $\mu_{r}$ does not change at all by applying this $SO(2)$ rotation. So this constitutes the isotropy subgroup. \newline
On the other hand dim$\big(T^{*}(Q_{int})\big)$ is by 6 smaller than dim$\big(T^{*}(Q_{cm})\big)$. Thus
\begin{equation}
dim\big(\frac{\mu^{-1}_{r}(a)}{G_{a}}\big) = dim\big(T^{*}(Q_{int})\big)+2.
\end{equation}  
So, in general the reduced phase space with respect to rotations $SO(3)$ is diffeomorphic to the cotangent bundle of the internal space $T^{*}(Q_{int})$. From the disscussion above, it is clear that the total group $SO(3)$ becomes an isotropy subgroup $G_{\textbf{a}}$ if and only if $\textbf{a}=0$, and in this case one has
\begin{equation}
\frac{\mu^{-1}_{r}(0)}{SO(3)}\cong T^{*}(Q_{int}).
\end{equation} 
Generally speaking, the reduced phase space is diffeomorphic to the fiber product $T^{*}(\frac{Q}{G})\times_{f}(\frac{Q}{G_{a}})$ over the quotient $\frac{Q}{G}$, keeping in mind that $\frac{Q}{G_{a}}$ is naturally identified with the coadjoint orbit bundle $Q\times_{G}(\frac{G}{G_{a}})$ over $\frac{Q}{G}$ (see \cite{20},\cite{21}) \newline 

Now we want to study the symplectic form $\sigma_{\textbf{a}}$ on the reduced phase space $P_{\textbf{a}}=\frac{\mu^{-1}_{r}(\textbf{a})}{G_{\textbf{a}}}$. Since $\sigma_{\textbf{a}}$ is defined by \[\pi^{*}_{\mu}\sigma_{\mu}=i^{*}_{\mu}\sigma\] and $\sigma=d\theta$. we have furthermore $i^{*}_{\mu}d\theta=d(i^{*}_{\mu}\theta)$ in our case. Remember that the  maps used  are $\pi_{r}:\mu^{-1}(r)\rightarrow P_{r}$ and $i_{r}:\mu^{-1}(r)\rightarrow T^{*}(Q_{cm})$.\newline

For notational convenience we work in the following on the tangent bundle over configuration space which is isomorphic to the cotangent bundle. Through this isomorphism the tangent bundle can be endowed with a canonical symplectic form, which we denote by the same letter we used for the  cotangent bundle, i.e. 
\begin{equation}\label{ocf}
\theta_{(x,v)}=\sum m_{k}<v_{k},dx_{k}>=K(v,dx)
\end{equation}
where each tangent space is equipped with a scalar product given by 
\begin{equation}\label{kmetric}
K_{x}(u,v)=\sum m_{k}(u_{k}\mid v_{k})
\end{equation} 
for $u=(u_{1},...,u_{n})$ and $v=(v_{1},...,v_{N})$ of $T_{x}(Q)$.\newline
Define \[\omega^{D}_{x}:\textbf{so}(3)\rightarrow T_{x}(Q)\] dual to $\omega_{x}: T_{x}\rightarrow \textbf{so}(3)$ where the following isomorphisms has been taken into account $\textbf{so}(3)\cong\textbf{so}^{*}(3)$ and $T_{x}(Q)\cong T^{*}_{x}(Q)$. For $\textbf{a}\in \textbf{so}(3)$ and $v\in T_{x}(Q)$, $\omega^{D}_{x}$ is defined by 
\begin{equation}\label{eq4.45}
(\omega_{x}(v)\mid\textbf{a})=:K_{x}(v,\omega^{D}_{x}(\textbf{a})).
\end{equation}
One can prove that for any $v\in T_{x}(Q)$ the vector $v-\omega^{D}_{x}\mu_{r}(x,p)$ with $I^{-1}_{x}(p)=v$ is vibrational (horizontal). To this end, it suffices to show that $v-\omega^{D}_{x}\mu_{r}(x,p)$ and $R_{\xi}(x)$ are orthogonal for any $\xi \in \wedge^{2}\mathbb{R}^{d}$, in other words showing $K_{x}(R_{\xi},v-\omega^{D}_{x}\mu_{r}(x,p))=0$, for $\forall \xi \in \wedge^{2}\mathbb{R}^{3}$ (see \cite{5}). \newline\newline
Taking \[(x,v)\] as a coordinate system on $T(Q)$, the submanifold $\mu^{-1}_{r}(\textbf{a})$ is determined in $T(Q)$ by the condition $R_{-\sum m_{k}x_{k}\wedge v_{k}}=\textbf{a}$.
Let now \[w=v-\omega^{D}_{x}\mu_{r}(x,p)\] with $I^{-1}_{x}(p)=v$, then the pair \[(x,w)\] meets the condition $R_{-\sum m_{k}x_{k}\wedge w_{k}}=0$, so that it serves as coordinate system in $\mu^{-1}_{r}(0)$ under that condition. A coordinate system on $\mu^{-1}_{r}(\textbf{a})$ can then be given by the pair $(x,w+\omega^{D}_{x}\textbf{a})$. With this in mind we rewrite the canonical one-form $\theta$ at point $(x,v)\in T(Q_{cm})$, which obviously has $T_{(x,v)}(T(Q_{cm}))$ as its domain

\[\theta_{(x,v)} = K(v,dx)\]\[= K_{x}(w,dx)+K_{x}\big(\omega^{D}_{x}\mu_{r}(x,p),dx\big)\]\[= K_{x}(w,dx)+\big(\mu_{r}(x,p)\mid \omega_{x}\circ dx\big),\]
where in the last step we have used \eqref{eq4.45}. Consequently on $\mu^{-1}_{r}(\textbf{a})$ we have
\begin{equation}
i^{*}_{\textbf{a}}\theta_{(x,v)}=K_{x}(w,dx)+(\textbf{a}\mid\omega_{x}\circ dx)
\end{equation}
where $v=w+\omega^{D}_{x} \textbf{a}$. Thus the canonical two-form $d\theta$ restricts to $d(i^{*}_{\textbf{a}}\theta)$ on $\mu_{r}^{-1}(\textbf{a})$;
\begin{equation}\label{4.47}
d(i^{*}_{a}\theta)= d\big(K_{x}(w,dx)\big)+d(\textbf{a}\mid\omega\circ dx).
\end{equation}
In all these equations $dx$ should be viewed as a vector $[dx_{1},dx_{2},...,dx_{N-1}]$. So you can formally act on it by the connection form $\omega$ and then take exterior derivative and so on.

Since $w$ is horizontal i.e. $ w\in T_{x,hor}\cong T_{\pi(x)}(Q_{int})$ the first term on the right hand side of  \eqref{4.47} is invariant under the action of $SO(3)$, and hence in one-to-one correspondence with the canonical two-form on $T(Q_{int})\cong T^{*}(Q_{int})$. In contrast, the second term of the same side, depending on x, cannot project to a two-form on $Q_{int}$. In fact $(\textbf{a}\mid d\omega)$ is not horizontal (its value changes if we act with the group on it, or to be more precise acting by any member of $G/G_{\textbf{a}}$). We recall that the horizontal part of $d\omega$ is defined as the curvature form. 
\newline

Last but not least, we discuss how an invariant metric\footnote{Invariant under action of the structure group.} on the total space of fiber bundles, induces metrics on horizontal and vertical subspaces. In the context of molecular physics this is known as \textit{splitting of energy into vibrational and rotational parts} \cite{5}, as the total space $T(Q_{cm})$ is the tangent bundle over the center of mass configuration space of a molecule, and the group of $3$-dimensonal rotations being the structure group.\newline 
Recall the decomposition \[T_{x}(Q_{cm})=T_{x,rot}\oplus T_{x,hor}\] and the orthogonal projections \[P_{x}:T_{x}\rightarrow T_{x,rot}\] and \[H_{x}:=(1_{x}-P_{x}):T_{x}\rightarrow T_{x,hor}\] where $1_{x}$ denotes the identity element in $T_{x}(Q_{cm})$. With the help of the connection form $\omega_{x}:T_{x}(Q_{cm})\rightarrow \textbf{so}(d)$, one has the following orthogonal decomposition for any $v\in T_{x}(Q_{cm})$ 
\[\omega_{x}(v)=P_{x}(v) \Rightarrow v=P_{x}(v)+H_{x}(v)\]
Hence for any $v$ and $u \in T_{x}(Q_{cm})$ one has
\begin{equation}\label{4.48}
K_{x}(v,u)=K_{x}\big(P_{x}(v),P_{x}(u)\big) + K_{x}\big(H_{x}(v),H_{x}(u)\big).
\end{equation}
Now if we set $v=u$, we obtain the kinetic energy expressed as the sum of rotational and vibrational energies. However this does not mean that there is no coupling between the rotational and vibrational motions. The coupling rather manifests itself into the dynamics through the connection form $\omega$. \newline
We now focus on the second term on the r.h.s of \eqref{4.48}. Let $\pi$ be the natural projection of $Q_{cm}$ onto $Q_{int}$. Differentiation of that map $\pi_{*}:T(Q_{cm})\rightarrow T(Q_{int})$ restricted on $T_{x,hor}$ gives an isomorphism of $T_{x,hor}$ with $T_{\pi(x)}(Q_{int})$. Let $X$,$Y\in T_{m}(Q_{int})$ for some $m\in Q_{cm}$. Then, at every point $x$ with $\pi(x)=m$, one has  unique horizontal vectors $v$ and $u$ satisfying $\pi_{*}(v)=X$ and $\pi_{*}(u)=Y$. If the metric $K_{x}$ is $SO(3)$-invariant, i.e. $K_{gx}(gv,gu)=K_{x}(v,u)$ then the vibrational energy (second term of \eqref{4.48}) induces a Riemannian metric $B$ on $Q_{int}$ through
\begin{equation}
B_{m}(X,Y):=K_{x}(v,u)
\end{equation} 
One can easily verify that this definition is independent of the choice of $x$ with $\pi(x)=m$.\newline\newline
We now look at the restriction of the vibrational energy to the submanifold $\mu^{-1}_{\textbf{a}}$.  Using the coordinate \[w=v-\omega^{D}_{x}\mu_{a}(x,p)\] (with $I^{-1}_{x}(p)=v$, paring the vectors and covectors \eqref{pvco}) the vibrational energy is written as $k_{x}(w,w)$ with $R_{-\sum m_{k}x_{k}\wedge w_{k}}=0$, and is in one-to-one correspondence with the kinetic energy of the internal motion. \newline
Now we turn to the first term on the r.h.s. of \eqref{4.48}, the rotational energy. Considering the definition of the  inertia operator $A_{x}$ of the configuration $x$, as a linear operator in $\wedge^{2}\mathbb{R}^{3}$, one can calculate (see \cite{5}) \[K_{x}(P_{x}(v),P_{x}(v))=K_{x}\big(\omega_{x}(v),\omega_{x}(v)\big)\]\[=(R A^{-1}_{x}R^{-1}\mu_{r}(x,p)\mid \mu_{r}(x,p)).\]  
If the system's Hamiltonian is rotation invariant the angular momentum is conserved i.e. $\mu=\textbf{a}$, and the last expression becomes a function of just the space variables $x$, namely $(R A^{-1}_{x}R^{-1}\mu\mid \mu)$. This function is in fact invariant under $G_{\textbf{a}}$ (easily verifiable by using (\ref{properties of R A}d), and thus projects down to a function on the reduced phase space, which can be seen as centrifugal potential. \newline

Now all the necessary ingredients are available for the reduction of the Hamiltonian system with respect to $SO(3)$. Remembering the inclusion map and the projection map 
\begin{subequations}
\begin{align}
      i_{\textbf{a}}:\mu^{-1}(\textbf{a})\rightarrow P=T(Q_{cm})\\
       \pi_{\textbf{a}}: \mu^{-1}(\textbf{a})\rightarrow P_{\textbf{a}}=\frac{\mu^{-1}(\textbf{a})}{G_{\textbf{a}}}.
\end{align}
\end{subequations} 
The reduced phase space $\frac{\mu^{-1}(\textbf{a})}{G_{\textbf{a}}}$ carries the symplectic form $\sigma_{\textbf{a}}$ which, as discussed before, is related to the canonical form $d\theta$ through \[i^{*}_{\textbf{a}}d\theta=\pi^{*}_{\textbf{a}}\sigma_{\textbf{a}}.\] 
On $\frac{\mu^{-1}(\textbf{a})}{G_{\textbf{a}}}$ the reduced Hamiltonian $H_{\mu}$ is defined by \[H_{\textbf{a}}\circ \pi_{\textbf{a}}=H\circ i_{\textbf{a}}.\] 
Note, that in the equations above defining  the reduced symplectic form $\sigma_{\textbf{a}}$, and the reduced Hamiltonian $H_{\textbf{a}}$, the usage of $\pi_{\textbf{a}}^{-1}$ is avoided, because the projection map $\pi_{\textbf{a}}$ is obviously not invertible (it sends a whole fiber to a point on the reduced space). This form and the Hamiltonian are expressed in coordinates $(x,v)$ on $\mu^{-1}_{r}(\textbf{a})$ with $v=w+\omega^{D}_{x}\textbf{a}$
\begin{subequations}
\begin{align}\label{rhh}
      \pi^{*}_{\textbf{a}}\sigma_{\textbf{a}}=i^{*}_{\textbf{a}}d\theta=d\big(K(w,dx)\big) +d(\textbf{a}\mid \omega)\\ \label{rhhh}
       H_{\textbf{a}}\circ\pi_{\textbf{a}}=H\circ i_{\textbf{a}}=\frac{1}{2}K(w,w)
\end{align}
\end{subequations} 
\[ +\frac{1}{2} \big(R A^{-1}_{x}R^{-1}\mu_{r}(x,p)\mid \mu_{r}(x,p)\big)+ U \]
where $R_{-\sum m_{k}x_{k}\wedge w_{k}}=0$.\newline 
The r.h.s. of these equations are invariant under $G_{a}$, and hence can be thought of as quantities on the reduced phase space. The first expressions on the r.h.s. of  \eqref{rhh} and \eqref{rhhh} are in one-to-one correspondence with the "canonical two-from", and the "kinetic energy" on $T^{*}(Q_{int})\cong T(Q_{int})$ respectively. The second expressions on the r.h.s. of \eqref{rhh} and \eqref{rhhh} can be seen as the source of the "coriolis force" and the "centrifugal potential" of the system's projected motion on the internal space (so the internal motion) respectively.   
 
\textbf{Note}, that if $a=0$, the reduced phase space is diffeomorphic to the cotangent bundle $T^{*}(Q_{int})$ of the internal space $Q_{int}$, and the symplectic form $\sigma_{\textbf{a}}$ becomes the canonical two form on $T^{*}(Q_{int})$. The reduced Hamiltonian $H_{a}$ is then a sum of the kinetic energy of internal (horizontal) motions and the potential on $Q_{int}$ (which is exactly the same potential as the one up on $Q_{cm}$).
If the system's motion on absolute space is planar (so $Q\cong \mathbb{R}^{2N}$), and $\textbf{a}\neq 0$ the reduced phase space is still diffeomorphic to $T^{*}(Q_{int})$, but the symplectic from $\sigma_{\textbf{a}}$ is the canonical one plus a two-form which can be seen as a "magnetic field" on $Q_{int}$. The reduced Hamiltonian $H_{\textbf{a}}$ also becomes the sum of kinetic and potential energies plus a centrifugal potential. In both these cases the system's motion is internal (horizontal). That means that it can be described on $T^{*}(Q_{int})$ or in terms of internal coordinates and their conjugate momenta.\newline


\newpage
\section{Reduction with respect to the similarity group $Sim(3)$}
\subsection{Metrics on the internal and shape space}
 Following \cite{10} we next review how the kinetic metric on absolute configuration space induces a metric on the internal configuration space $Q_{int}=\frac{Q_{cm}}{SO(3)}$. Thereafter, we explain a new way to derive a metric on shape space from the mass metric on absolute configuration space in a unique way. \newline\newline
\textbf{Metric on the internal space:}\newline\newline
Let us recall how the metric on the internal space $Q_{int}=\frac{Q_{cm}}{SO(3)}$ was derived from the $SO(3)$-invariant mass metric \eqref{kmetric} on the  center of mass system
\begin{subequations}
\begin{align}
       \textbf{M}_{x}(u,v)=\sum m_{k}<u_{k}\mid v_{k}>\\
       \textbf{M}_{x}(u,v)=\textbf{M}_{gx}(gu,gv)
\end{align}
\end{subequations}
where $u=(u_{1},...,u_{n})$ and $v=(v_{1},...,v_{N}) \in T_{x}(Q_{cm})$ being any two tangent vectors of $Q_{cm}$ at the point $x\in Q_{cm}$. 

Now given two internal vectors $v',u'\in T_{q}(Q_{int})$, there are unique vectors $u,v\in T_{x}(Q_{cm})$ \footnote{namely their horizontal lifts (\ref{hli}).} so that 
\[\left\{
  \begin{array}{lr}
    \pi(x)=q\\
    \pi_{*}(u)=u'\\
    \pi_{*}(v)=v'.
  \end{array}
\right.
\] 
Now the metric $B$ on $Q_{int}$ can be defined in the following way
\begin{equation}
B_{q}(v',u'):=\textbf{M}_{x}(v,u).
\end{equation}
Since the metric $M$ is $SO(3)$-invariant, it does  not make any difference to which $x\in \pi^{-1}(q)$ the internal vectors $v,u$ had been lifted  for the value assigned by $B_{q}$.  This is, in fact, crucial for the well-definedness of the reduced metric. \newline \newline
The kinetic energy of a $N$-particle system in the center of mass frame is $K=\frac{1}{2}\sum_{\alpha=1}^{N-1}\mid\dot{\pmb{r}}_{\alpha}\mid^{2}$. Using $\dot{\pmb{r}}_{b\alpha}=\pmb{\omega}\times \pmb{r}_{b\alpha}+\frac{\partial\pmb{r}_{b\alpha}}{\partial q^{\mu}}\dot{q}^{\mu}$, and the expression  \[\pmb{A}_{\mu}(q)=I^{-1}\pmb{a}_{\mu}\] for the gauge potentials, where
\begin{equation}\label{gaugepo}
  \pmb{a}_{\mu}=\pmb{a}_{\mu}(q):=\sum_{\alpha=1}^{N-1}\pmb{r}_{b\alpha}\times\frac{\partial\pmb{r}_{b\alpha}}{\partial q^{\mu}}  
\end{equation}
and $A$ being the moment of inertia tensor with components \[A_{ij}=A_{ij}(q):=\sum_{\alpha=1}^{N-1}(\mid\pmb{r}_{b\alpha}\mid^{2}\delta_{ij}-r_{b\alpha i}r_{b\alpha j})\] 
one can write down the kinetic energy as 
\begin{equation}\label{lkin}
K=\frac{1}{2}<\pmb{\omega}\mid A\mid\pmb{\omega}>+<\pmb{\omega}\mid A\mid\pmb{A}_{\mu}>\dot{q}^{\mu}
\end{equation}
\[+\frac{1}{2}h_{\mu\nu}\dot{q}^{\mu}\dot{q}^{v}\]
with 
\begin{equation}
h_{\mu\nu}=h_{\mu\nu}(q)=\sum_{\alpha=1}^{N-1}\frac{\partial\pmb{\rho}_{\alpha}}{\partial q^{\mu}}.\frac{\partial\pmb{\rho}_{\alpha}}{\partial q^{\nu}}.
\end{equation}
The velocity of a system's configuration in Jacobi coordinates is given by a vector \[\mid v>=[\dot{\pmb{r}}_{s1},...,\dot{\pmb{r}}_{s,n-1}]\] and in orientational and internal coordinates by the vector \[\mid v>=[\dot{\theta}^{i},\dot{q}^{\mu}]\] where $\theta^{i}$ are the Euler angles which turn the space frame to the body frame of a configuration. If one decides to use the components of the body angular velocity $\pmb{\omega}$ instead of the time derivatives of the Euler angles for denoting vectors in $T(SO(3))$ the configuration's velocity can alternatively be expressed as \[\mid v>=[\pmb{\omega},\dot{q}^{\mu}]\] in angular velocity and internal basis. This last combination forms an anholonomic frame or vielbein on $T(Q_{cm})$. Remember the relation between the body components of angular velocity and derivatives of Euler angles
\[ \begin{bmatrix}\omega_{1} \\ \omega_{2} \\ \omega_{3} \end{bmatrix}=\begin{bmatrix}-sin\beta cos\gamma & sin\gamma & 0 \\ sin\beta sin\gamma & cos\gamma & 0 \\ cos\beta & 0 & 1 \end{bmatrix} \begin{bmatrix} \dot{\alpha} \\ \dot{\beta}\\ \dot{\gamma} \end{bmatrix}.\]
So the (kinetic)metric tensor $m_{Q_{cm}}$ in angular and internal basis vectors $\{\omega^{i},\dot{q}^{\mu}\}$, where $i=1,2,3$, and $q=1,...,3N-6$ becomes
\[
<v\mid v>=\begin{bmatrix}\pmb{\omega}^{T} & \dot{q}^{\mu} \end{bmatrix}\begin{bmatrix}A & A\pmb{A}_{\nu} \\ \pmb{A}_{\mu}^{T}A & h_{\mu\nu}\end{bmatrix}\begin{bmatrix}\pmb{\omega} \\ \dot{q}^{\nu} \end{bmatrix}
\]
\[ =(m_{Q_{cm}})_{ab}v^{a}v^{b} .\]
So, the metric on $Q_{cm}$ in angular and internal basis vectors $[\pmb{\omega},\dot{q}^{\mu}]$ is given by 
\begin{equation}\label{mmavsv}
    \textbf{M}_{ab}=\begin{bmatrix}A & A\pmb{A}_{\nu} \\ \pmb{A}_{\mu}^{T}A & h_{\mu\nu}\end{bmatrix}.
\end{equation}
Decomposition of an arbitrary system's velocity in horizontal and vertical parts gives
\[\mid v>=\mid v_{v}>+\mid v_{h}>\]
\[[\pmb{\omega},\dot{q}^{\mu}]=[\pmb{\omega}+\pmb{A}_{v}\dot{q}^{v},0]+[-\pmb{A}_{\nu}\dot{q}^{\nu},\dot{q}^{\mu}] \]
Correspondingly, the kinetic energy of the system can  also  be thought of as the addition of two separate vertical and horizontal kinetic energies, i.e.
\[K=K_{v}+K_{h}=\frac{1}{2}(\pmb{\omega}+\pmb{A}_{\mu}\dot{q}^{\mu})A(\pmb{\omega}+\pmb{A}_{\nu}\dot{q}^{\nu})\]\[+\frac{1}{2}B_{\mu\nu}\dot{q}^{\mu}\dot{q}^{v}\]
where $B_{\mu\nu}$ is the metric on internal space
\[B_{\mu\nu}=h_{\mu\nu}-\pmb{A}_{\mu}A\pmb{A}_{\nu}.\]
So, in summary, to a vector \[\mid v'>=\dot{q}^{\mu}\] on internal space $Q_{int}$ we associate a vector $\mid v_{h}>$ on $Q_{cm}$ which is called its \textit{horizontal lift}, connecting the two fibers. In the basis made of angular and shape velocities, the horizontal lift of $v'$ takes the from
\begin{equation}\label{hli}
    \mid v_{h}>=[-\pmb{A}_{\mu}\dot{q}^{\mu},\dot{q}^{\mu}].
\end{equation}
Then, the metric $B_{\mu\nu}$ on the internal space can be found by the following defining equation 
\[<v'_{1}\mid v'_{2}>=B_{\mu\nu}\dot{q}_{1}^{\mu}\dot{q}_{2}^{\nu}:=<v_{1h}\mid v_{2h}>\]\[=\textbf{M}_{ab}v^{a}_{1h}v^{b}_{2h}\] which leads to
\begin{equation}\label{internalmetricmanybody}
   B_{\mu\nu}=h_{\mu\nu}-\pmb{A}_{\mu}A\pmb{A}_{\nu}.
\end{equation}
For more information about the derivation of the metric on internal space, we highly recommend \cite{10}. \newline\newline
\textbf{Metrics on shape space:}\newline\newline
Now we are ready to derive a metric $N$ on shape space $S=\frac{Q}{sim(3)}$. Since the mass metric $M$ is not scale invariant, it is generally believed that, in contrary to $Q_{int}$, it does not directly induce a metric on  shape space $S$.  Once one takes the change of the ``flow of time'' (measured in Planck units) under scale transformation into account  , one can see that the original mass metric $M_{x}$ directly induces a unique metric on Shape space. We first review how the metric on shape space is derived with introduction of a conformal factor, and then give our new derivation of the corresponding metric on shape space.\newline\newline
As is explained in \cite{2} one can introduce a new $sim(3)$-invariant metric on $Q$, which subsequently  induces a metric on shape space in a natural way. As the mass metric $\textbf{M}$ is already rotation- and translation-invariant, the easiest way to arrive at a similarity-invariant metric is to multiply the mass metric by a function $f(x)$ (the so called conformal factor), so that the full expression \[\textbf{M}'_{x}:=f(x)\textbf{M}_{x}\]   becomes scale invariant, i.e. 
\[\forall c\in \mathbb{R}^{+}, \forall u,v \in T_{x}(Q) :\]\[f(c x)\textbf{M}_{c x}(c u,c v)= f(x)\textbf{M}_{x}(u,v).\]
Note that the function $f$ must itself be translation- and rotation-invariant so that it doesn't spoil the nice properties of the mass metric. As $M'_{x}=f(x)M_{x}$ is now a metric invariant under the full similarity group we are ready to write down the metric $N$ on shape space:
\begin{equation}\label{shapemetric}
N_{s}(v',u'):=\textbf{M}'_{x}(v,u)=f(x)\textbf{M}_{x}(v,u),
\end{equation}         
where
\[\left\{
  \begin{array}{lr}
    \pi(x)=s\\
    \pi_{*}(u)=u'\\
    \pi_{*}(v)=v'
  \end{array}
\right.
\] 
with the projection map $\pi:Q_{cm}\rightarrow S=\frac{Q}{sim(3)}$.\newline
From the behavior of the mass metric $\textbf{M}$ under dilations i.e. $\textbf{M}_{c x}(c u,c v)=c^{2} \textbf{M}_{c x}(v,u)$ one easily sees that any rotation- and translation-invariant homogeneous function\footnote{ A function of $r$ variables $x_{1},...,x_{r}$ is being called homogeneous of degree $n$ if $f(c x_{1},...,c x_{r})=c^{n}f(x_{1},...,x_{r})$,$\forall c$ } of degree $-2$  perfectly meets all the requirements of a conformal factor. For instance
\begin{equation}\label{cf1}
    f(x)=\sum_{i<j}\mid\mid \pmb{x}_{i}-\pmb{x}_{j}\mid\mid^{-2}
\end{equation}
or 
\begin{equation}\label{cf2}
    f(x)=I^{-1}_{cm}
\end{equation}
 where 
 \[I_{cm}(x)=\sum_{j}m_{j}\mid\mid \pmb{x}_{j}-\pmb{x}_{cm}\mid\mid^{2}\]\[=\frac{1}{\sum_{i}m_{i}}\sum_{i<j}m_{i}m_{j}\mid\mid \pmb{x}_{j}-\pmb{x}_{i}\mid\mid^{2}\] 
 are two legitimate examples of conformal factors (as suggested in \cite{2}). However, as the introduction of conformal factors  leads to  the appearance of unphysical forces (see appendix i for clarification), and as there exists some arbitrariness in the choice of this function, we prefer to avoid this method here. Below we propose another way to derive the corresponding metric on shape space which does not show the problems just mentioned.\newline

Bearing in mind that measurements of the velocity are something experimental, the transformation law of the velocities under scale transformations of the system (or any other transformation of the system) must also include experimental reasoning. Based on the principle of relationalism we showed that the behavior of rods and clocks under scale transformations of the system is such that the measured velocities of objects (or parts of the system) are invariant. This is a natural consequence of the simultaneous expansion of the measuring rod and a corresponding dilation of unit of time (See section II.iv for an explanation of this fact).\newline
Hence a velocity vector \[v_{x}=(v_{1},...,v_{N})\in T_{x}(Q_{cm})\] of an N-particle system transforms under dilatations of the system as \[x\rightarrow c x\]
\[v_{x}=(v_{1},...,v_{N})\in T_{x}(Q_{cm})\]\[\rightarrow v_{c x}=(v_{1},...,v_{N})\in T_{c x}(Q_{cm}).\]  
Hence for this transformation of velocities under scale transformations of the system, the mass metric is indeed scale invariant in a trivial way, i.e. \[\textbf{M}_{x}(v_{x},u_{x})\rightarrow \textbf{M}_{c x}(v_{c x},u_{c x})= \textbf{M}_{x}(v_{x},u_{x}).\]
In other words, the mass metric $M$ expressed in special coordinates on $Q_{cm}$ built from internal rods (for example Planck's unit of length), is scale invariant.

So the metric $N$ of the shape space can be written as
\begin{equation}\label{shapemetric1}
N_{s}(v',u'):=\textbf{M}_{x}(v,u),
\end{equation}         
where
\[\left\{
  \begin{array}{lr}
    \pi(x)=s\\
    \pi_{*}(u)=u'\\
    \pi_{*}(v)=v'
  \end{array}
\right.
\] 
with the projection map $\pi:Q_{cm}\rightarrow S=\frac{Q}{sim(3)}$. The metric $N$ is defined as before and it  does not make any difference for the value assigned by $N_{s}$ to which $x\in \pi^{-1}(q)$ the shape vectors $v',u'$ are lifted, because the metric $\textbf{M}$ is $Sim(3)$-invariant

\subsection{Reduction of the theory} 

For the purpose of reduction of classical mechanics w.r.t. scale transformations, we now use the methods explained in chapter III. One of the reasons that so far nobody  has gone after extension of these formalisms to the similarity group is obviously that the potential function defined on  absolute space, though being manifestly rotational and translational invariant, is clearly not scale invariant (take Newtonian gravity as an example). However as explained in chapter II, scale transformation becomes an additional symmetry of classical physics (see equation \eqref{spo}) and this enables us to perform a symplectic reduction of the system's phase space with respect to the whole similarity group $G=sim(3)$.\newline   

Beside having a similarity invariant potential function on absolute configuration space (see equation \eqref{spo}), in order to reduce the classical systems with respect to the similarity group (which was originally argued for and motivate in chapter II) with the help of symplectic reduction methods explained in chapter III, we have to change the connection form (\ref{rotconnfor1}) to the following one\newline
\[\omega=\omega_{r}+\omega_{s} \]
\begin{equation}\label{shapeconnection}
=R\bigg(A_{x}^{-1}(\sum_{j=1}^{N-1}r_{j}\times dr_{j})\bigg)+I_{3}\textbf{D}_{x}^{-1}\bigg(\sum_{j=1}^{N-1} r_{j}.dr_{j}\bigg)
\end{equation}
in which $I_{3}$ is the $3\times 3$ identity matrix, and we have defined the operator \[
\textbf{D}_{x}:\mathbb{R}\rightarrow\mathbb{R}\]
\[\dot{\pmb{\lambda}}\rightarrow D \]
\[\textit{expansion velocity } \rightarrow \textit{dilational momentum }  \]
as 
\begin{equation}\label{dilmome}
\textbf{D}_{x}(\dot{\pmb{\lambda}}):=\sum_{j=1}^{N-1}r_{j}^{2}\dot{\pmb{\lambda}}
\end{equation} 
and we call it the "\textit{dilational tensor}". Here $\dot{\pmb{\lambda}}$ stands for the rate of change of scale of the system (scale velocity so to speak)\footnote{Here of course we assume that all measurements are conducted with the use of special Newtonian rods and clocks, which are isolated from the materialized universe and do not get affected by them in any way or by transformations we perform on the materialized universe. Practically of course such measuring instruments do not exist, but existence of absolute space and absolute time in Newtonian world view justifies their hypothetical existence.}
\begin{equation}
\dot{\pmb{\lambda}}:=\frac{\dot{\lambda}}{\lambda}
\end{equation}
with 
\begin{equation}\label{scvari} 
\lambda:=max\mid\pmb{x}_{i}-\pmb{x}_{j}\mid
\end{equation}
for $i,j=1,...,N$ being the system's scale variable.

We constructed this operator in direct analogy to the inertia tensor $A_{x}$. The inertia tensor sends an angular velocity (which can be represented as a vector in $\mathbb{R}^{3}$) to another vector in $\mathbb{R}^{3}$ which represents the total angular momentum of the whole system (object). In the same way, the dilational tensor $\textbf{D}_{x}$ takes an expansion velocity, which can be represented by just a number in $\mathbb{R}$ to a measure of total expansion of the system (dilational momentum $D$) which again can be represented by another number in $\mathbb{R}$. As the Lie algebra of the group $\frac{sim(3)}{trans(3)}$ can be considered to be the addition of $3\times 3$ skew-symmetric matrices of $\textbf{so}(3)$, and real multiples of $3\times 3$ identity matrix $I_{3}$, one recognizes the correct structure in this connection form. 
If one takes a random vector of $T_{x}(M)$ and act on it by this connection form, the first term of  \eqref{shapeconnection} gives a member of $\textbf{so}(3)$, and the second term, a number multiplied by the identity matrix $I_{3}$. So, it does what it is expected to do.

\subsubsection{Symplectic reduction of  phase space}
Now that the metric \eqref{shapemetric1} on  shape space, and the connection form \eqref{shapeconnection} is given, the way to get the reduced Hamiltonian equations of motion with respect to the similarity group is paved.

The first step is to find the momentum mapping corresponding to the group \[G=SO(3)\times\mathbb{R}^{+}.\] To this end consider first the Lie algebra of  $G$. It can be written as \[\textbf{G}= z I_{3}+\textbf{so}(3)\] where $z\in \mathbb{R}$ and $\textbf{so}(3)$ are as usual the skew-symmetric $3\times 3$ matrices standing for the lie algebra of the rotations group.  \newline
The action of $G$ on $T^{*}(Q_{cm})$ is the following \newline

\begin{equation}\label{actsim}
(x_{1},...,x_{N-1};p_{1},...,p_{N-1})
\end{equation}
\[\downarrow \]
\[ (c g x_{1},...,c g x_{N-1};  g p_{1},..., g p_{N-1}), \]
where, as before, $c\in R^{+}$ and $g\in SO(3)$ (and all the lengths and time intervals are measured by the absolute Newtonian rods and clocks.).

The momentum mapping corresponding to  scale transformations is $\mu_{scale}=\sum_{j=1}^{N-1} r_{j}.dr_{j}$. Hence the momentum mapping for the group \footnote{of both rotations and scale transformations} $G$ becomes
\begin{equation}
\mu_{sim}=I_{3}\sum_{j=1}^{N-1} r_{j}.dr_{j}+\sum_{j=1}^{N-1}R_{-r_{j}\wedge p_{j}}.
\end{equation}
We call $\mu_{sim}$ the similarity momentum \footnote{with slight abuse of notation, since starting with the center of mass system translations are not taken into account.}. Remember that mathematically the similarity momentum is the following mapping 
\begin{equation}
\mu_{sim}:P=T^{*}(Q_{cm})\rightarrow \mathbb{R}I_{3}+\textbf{so}(3):=\textbf{G}\cong \textbf{G}^{*}.
\end{equation}
In fact $\textbf{G}$ is the summation of a skewsymmeric matrix with a real multiple of the identity matrix.

Let $a\in \textbf{G}\cong \textbf{G}^{*}$. It can be rewritten as $a=DI_{3}+\textbf{L}$ for some $D\in \mathbb{R}$ standing for dilational momentum (diagonal part of $a$), and some $\textbf{L}\in \textbf{so}(3)$ standing for the angular momentum (the non-diagonal part of $a$). 
As before, $\mu_{G}^{-1}(a)$ is a submanifold of $P=T^{*}(Q_{cm})$. The reduced phase space $\frac{\mu_{G}^{-1}(a)}{G_{a}}$ is then achieved by quotienting $\mu_{G}^{-1}(a)$ with respect to the isotropy group $G_{a}$.\newline 
The isotropy group corresponding to  rotations was already discussed in chapter $3$. Now it's time to discuss the same question for the scale transformations. For simplicity consider a system which is purely expanding. The dilational momentum of this system is \[D=\sum_{i=1}\textbf{r}_{i}.\textbf{v}_{i}.\] 
Now perform a scale transformation $\textbf{r}_{i}\rightarrow c \textbf{r}_{i}$. Having the corresponding transformation law of the velocities \eqref{actsim} in mind, the dilational momentum transforms as 
\begin{equation}\label{sssc}
   D\rightarrow D'=\sum_{i}(c\textbf{r}_{i}).\textbf{v}_{i}=c D .
\end{equation}
This should be compared to the system's angular momentum, which  is also in general\footnote{whenever the system's angular momentum is non-vanishing} not invariant under rotations\footnote{Note however, using instead of  absolute Newtonian rods and clocks internal rods and clocks, dilational momentum becomes invariant under scale transformations.}.  
From \eqref{sssc} it becomes clear that any scale transformation changes the value of the dilational momentum, except if $D=0$. Hence the isotropy subgroup of the scale transformations is $\oslash$ when $D\neq 0$, and $\mathbb{R}^{+}$ when $D=0$. 

Now we are in a position to compare the reduced phase space $\frac{\mu_{G}^{-1}(a)}{G_{a}}$  with respect to $G=\mathbb{R}^{+}\circ SO(3)$ with the cotangent bundle of  shape space $T^{*}(S)$. Remember that $dim\big(T^{*}(Q_{cm})\big)=6N-6$ and $dim\big((T^{*}(S)\big)=6N-14$. Thus when going form the cotangent bundle of the center of mass system $T^{*}(Q_{cm})$ to the cotangent bundle of the shape space $T^{*}(S)$,  $8$ dimensions get eliminated.

Choosing a specific value $a\in\textbf{G}$ for the similarity momentum of the system reduces the dimension of $T^{*}(Q_{cm})$ by $4$. Consequently taking the quotient with respect to the corresponding isotropy group $G_{a}$, leads to an elimination of $1$ extra dimension when $\textbf{L}\neq 0$ and $D\neq 0$ (because in this case $G_{a}=SO(2)$ ) and $4$ extra dimension when $\textbf{L}= D= 0$. For this latter case $8$ dimensions are eliminated in total and hence the reduced phase space becomes isomorphic to the cotangent bundle of  shape space i.e. \[\frac{\mu_{G}^{-1}(0)}{G_{0}}\cong T^{*}(S).\] 
For the generic case of $\textbf{L}, D\neq 0$ we have 
\begin{equation}
dim\big(\frac{\mu_{G}^{-1}(a)}{G_{a}}\big)= dim\big(T^{*}(S)\big)+3.
\end{equation}       
Now we wish to discuss the canonical form and its reduction with respect to $G=SO(3)\times \mathbb{R}^{+}$. In particular we seek the reduced symplectic form $\sigma_{a}$ on the reduced phase space \[P_{a}=\frac{\mu_{r}^{-1}(a)}{G_{a}}\] starting as usual with the canonical one-form $\theta=\sum m_{k}<v_{k},dx_{k}>=K(v,dx)$ as we saw in \eqref{ocf}. Remember that $\sigma_{a}$ was defined by $\pi^{*}_{\mu}\sigma_{\mu}=i_{\mu}^{*}\sigma$ with $\sigma=d\theta$ and the inclusion map $i_{\mu}$ and the projection map $\pi_{a}$ were defined as the following 
\begin{subequations}
\begin{align}
      i_{a}:\mu^{-1}_{sim}(a)\rightarrow P=T^{*}(Q_{cm})\\
       \pi_{a}: \mu^{-1}_{sim}(a)\rightarrow P_{a}=\frac{\mu^{-1}_{sim}(a)}{G_{a}}.
\end{align}
\end{subequations} 
Define an operator dual to the connection form \eqref{shapeconnection} as \[\omega_{x}^{D}:\textbf{G}=\mathbb{R}I_{3}+\textbf{so}(3)\rightarrow T_{x}(Q_{cm}).\] 
For $a\in \textbf{G}$ and $v\in T_{x}(Q_{cm})$, $\omega_{x}^{D}$ was defined in \eqref{eq4.45}. We mentioned in section IV.ii that for any $v\in T_{x}(Q)$ the vector $v-\omega^{D}_{x}\mu_{sim}(x,p)$ with $I^{-1}_{x}(p)=v$ is horizontal. \newline
 We choose $(x,v)$ as coordinate system in $T(Q_{cm})$. The submanifold $\mu^{-1}_{sim}(a)$ is determined in $T(Q_{cm})$ by the condition \[I_{3}\sum_{j=1}^{N-1}x_{j}.v_{j}+\sum_{j=1}^{N-1}R_{-m_{j}x_{j}\wedge v_{j}}=a.\]
Now with the help of $\omega^{D}$ decompose a vector $v\in T(Q_{cm})\cong T^{*}(Q_{cm})$ into horizontal and vertical part, i.e. $v=w+\omega_{x}^{D}\mu_{sim}(x,p)$ where again $I^{-1}_{x}(p)=v$. Rewrite the canonical one-form $\theta$ as the following
\[\theta_{(x,v)}=\textbf{M}(v,dx)=\textbf{M}\big(w+\omega_{x}^{D}\mu_{sim}(x,p),dx\big)\]\[=\textbf{M}(w,dx)+K\big(\omega_{x}^{D}\mu_{sim}(x,p),dx\big)\]\[=\textbf{M}(w,dx)+\big(\mu_{sim}(x,p)\mid \omega_{x}\circ dx\big).\]
Hence on $\mu_{sim}^{-1}(a)$ we get the following one-form
\begin{equation}
i^{*}_{a}\theta_{(x,v)}=\textbf{M}(w,dx)+(a\mid\omega_{x}\circ dx),
\end{equation}
where we again used the horizontal-vertical decomposition of vectors $v=w+\omega_{x}^{D}a$. \newline
The canonical two-from on $\mu_{sim}^{-1}(a)$ then becomes
\begin{equation}\label{rew}
d(i^{*}_{a}\theta)=d\big(\textbf{M}(w,dx)\big)+d(a\mid \omega\circ dx).
\end{equation}
Since $w$ is horizontal i.e. $ w\in W_{x,hor}\cong T_{\pi(x)}(S)$ the first term in the right hand side of  \eqref{rew} is invariant under $G=\mathbb{R}^{+}\times SO(3)$, and hence in one-to-one correspondence with the canonical two-form on $T(S)\cong T^{*}(S)$. Contrary to this the second term of the same side, depends on x, and hence does not project to any two-form on $T^{*}(S)$. As a matter of fact $(a\mid d\omega)$ is not horizontal (its value changes if we act with the group on it, or to be more precise acting by any member of $G/G_{a}$).\newline 

Having the following substitutions for the similarity momentum \[a=DI_{3}+\textbf{L}\] and for the connection form \eqref{shapeconnection}, \[\omega=\omega_{r}+\omega_{s}\] where as before \[\omega_{r}:T(Q_{cm})\rightarrow \textbf{so}(3)\] and \[\omega_{s}:T(Q_{cm})\rightarrow \mathbb{R}I_{3}\] 
we rewrite the second term in \eqref{rew} as follows
\[d(a\mid \omega\circ dx)=d\big(DI_{3}+\textbf{L}\mid(\omega_{r}+\omega_{s})\circ dx\big)\]\[=d(\textbf{L}\mid \omega_{r}\circ dx)+d(DI_{3}\mid \omega_{s}\circ dx).\]
 Here we also used the fact that a diagonal and a skew-symmetric matrix are perpendicular to each other w.r.t. the matrix inner product $(A\mid B)=\frac{1}{2}tr(AB^{T})$. Hence, the two-form on  reduced phase space $\mu_{sim}^{-1}(a)$ can be rewritten:
\begin{equation}\label{rew1}
d(i^{*}_{a}\theta)=d\big(\textbf{M}(w,dx)\big)+d(\textbf{L}\mid \omega_{r}\circ dx)
\end{equation}
\[+d(DI_{3}\mid \omega_{s}\circ dx).\]
As mentioned before in general (\textbf{L}$\neq 0$ and $D\neq 0$) neither the second term nor the third term is in one-to-one correspondence with two-forms on $T^{*}(S)$ as a result of the non-trivial transformation laws of $\textbf{L}$ and $D$ under the group $\frac{G}{G_{a}}$. However, as these terms are invariant under $G_{a}$, they can legitimately be considered as differential forms on   reduced phase space $P_{a}$. The second term is the source of \textit{Coriolis force} \cite{5}.  We call the last term \textit{dilational force}.   Note that even when the overall angular momentum of the system under consideration  vanishes, i.e. \textbf{L}$=0$, its overall expansion (or contraction) causes extra complication for the reduced motion on $S$, so that it fails to be a describable in $T^{*}(S)$. So the dilational force, like the Coriolis force, is a non-Hamiltonian force (cannot be derived from a potential function on $S$). \newline     

A similar procedure as the one seen in the previous section leads to a splitting of the kinetic energy. Having the decomposition \[T_{x}(Q_{cm})=T_{x,sim}\oplus T_{x,hor}\] and the orthogonal projections \[P_{x}=T_{x}\rightarrow T_{x,sim}\] and \[H_{x}=1_{x}-P_{x}\] in mind, for any $v$ and $u\in T_{x}(Q_{cm})$ one has \[\textbf{M}_{x}(v,u)=\]\[\textbf{M}_{x}\big(P_{x}(v),P_{x}(u)\big)+\textbf{M}_{x}\big(H_{x}(v),H_{x}(u)\big).\] 
Now if we set $v=u$, we obtain the kinetic energy expressed as the sum of similarity and shape energies. In particular the similarity energy can be written down as 
\begin{equation}\label{spl}
\textbf{M}_{x}\big(P_{x}(v),P_{x}(v)\big)=\big(RA_{x}^{-1}R^{-1}\textbf{L}\mid\textbf{L}\big)
\end{equation}
\[+\big(I_{3}\textbf{D}^{-1}D\mid I_{3}D\big),\]
where $R$ and $A_{x}$ were defined as
\[R: \mathbb{R}^{3}\rightarrow \textbf{so(3)} \]
\[
R(a)=\begin{bmatrix}0 & -a^{3} & a^{2} \\a^{3} & 0 & -a^{1}\\ -a^{2} & a^{1} & 0 \end{bmatrix}
\]
for $a\in \mathbb{R}^{3}$, as explained in more extend in appendix iv, and
\[A_{x}: \mathbb{R}^{3}\rightarrow \mathbb{R}^{3}\]
\[
    A_{x}(\pmb{v})=\sum_{j=1}^{N-1}\pmb{r}_{j}\times(\pmb{v}\times \pmb{r}_{j}),   \pmb{v}\in\mathbb{R}^{3}, x\in Q_{cmns}.
\]
The symplectic form $\sigma_{a}$ on the reduced phase space $P_{a}=\frac{\mu_{sim}^{-1}(a)}{G_{a}}$ can as before be derived from the canonical 2-from $d\theta$ through \[i^{*}_{a}d\theta=\pi^{*}_{a}\sigma_{a}\] and the reduced Hamiltonian $H_{a}$ on $P_{a}$ is defined through $H_{a}\circ\pi_{a}=H\circ i_{a}$.\footnote{As before the inclusion map was defined as \[i_{a}:\mu^{-1}_{sim}(a)\rightarrow P=T^{*}(Q_{cm})\]} Choosing the coordinate system $(x,v)$ on $\mu_{sim}^{-1}(a)$, where $v=\omega^{D}(a)$, one gets the following expressions for the reduced two-form, and reduced Hamiltonian
\begin{equation}\label{rshh1}
      \pi^{*}_{a}\sigma_{a}=i^{*}_{a}d\theta=d\big(\textbf{M}(w,dx)\big)+d(\textbf{L}\mid \omega_{r}) 
      \end{equation}
      \[ +d(DI_{3}\mid \omega_{s})\]
      \begin{equation}\label{rshh2}
          H_{a}\circ\pi_{a}=H\circ i_{a}=\frac{1}{2}\textbf{M}(w,w)+\frac{1}{2} \big(RA_{x}^{-1}R^{-1}\textbf{L}\mid\textbf{L}\big)
      \end{equation}\[+\frac{1}{2}\big(I_{3}\textbf{D}^{-1}D\mid I_{3}D\big)+ U .\]
The last term at the right hand side of (\ref{rshh1}) can be understood as a new kind of fictitious force, which we call \textit{dilational force}. It has a  similar nature as the Coriolis force. The third term on the right hand side of (\ref{rshh2}) can also be understood as a new kind of potential, which we call \textit{scale potential}. The scale force, which is defined as the gradient of the scale potential, is in its nature similar to the centrifugal force.   
\newpage
\section{Conclusions}
In this paper we reviewed the method of symplectic reduction of a dynamical system with respect to a symmetry group. To illustrate this method, we also reviewed its application in reducing classical mechanics with respect to the Euclidean group $E(3)$.\newline
We introduced a new physical principle, i.e. the "\textit{Principle of Relationalism}", as a guideline for the implementation of relational ideas in modern physics. This principle enabled us to make  classical physics scale invariant, which in turn ensured the existence of laws of motion on shape space. So it enabled us to achieve a full relational reading of classical physics, which was hiding itself behind some foundational incompletenesses of the classical theories, i.e. lack of any derivation for the so called "constants of nature" from within the theories.  \newline
We then performed the symplectic reduction of the new scale invariant theory on absolute configuration space, and derived the reduced equations of motion (for the system's shape degrees of freedom). The appearance of two new forces on the reduced phase space were noticeable.

 \newpage
\section{Appendix}
\subsection{Geodesics as true evolution on shape space?}
One can assume\footnote{Arguments for this assumption are usually based on simplicity and beauty which are being seen in reducing all of physics to geometry as strongly promoted by Descartes.} that the true evolution on shape space is exactly along the geodesics corresponding to the metric \eqref{shapemetric} on shape space, as is the case in \cite{1} and \cite{2}. Of course,  horizontal lifting of any geodesic from shape space $S$ to  absolute configuration space $Q$
also results in a geodesic with respect to a conformal metric $\textbf{M}'$ there. Now consider the Jacobi action on $Q$ 
\begin{equation}
\bar{S}=\int_{\tau_{A}}^{\tau_{B}}\sqrt{(E-V)}\frac{dl}{d\tau}d\tau=\int_{\tau_{A}}^{\tau_{B}}\sqrt{(E-V)}dl
\end{equation}
and take $E=0$. To ensure that the corresponding action principle $\bar{S}=\int_{\tau_{A}}^{\tau_{B}}\sqrt{-V}dl$ \footnote{$dl^{2}=g_{ij}dx^{i}dx^{j}$} always leads to a geodesics on absolute configuration space, $V$ must indeed take the role of a conformal factor, i.e. $V=-f(x)$. Thus  the expression $\sqrt{-V}dl$ becomes the line element with respect to the new metric $\textbf{M}'$, as is elaborated in \cite{2}. So, using (\ref{cf1}) for the conformal factor (as one among many options) one gets $V=\sum_{i<j}\frac{1}{\mid\mid \pmb{x}_{i}-\pmb{x}_{j}\mid\mid^{2}}$, where the summation goes over all particles. The characteristic feature of this potential is its homogeneity of degree $-2$, which is a necessary condition for a conformal factor.    
As the main classical interaction potentials between particles are known to be inversely proportional to their distance from each other, the previous result seems at first glance to be empirically inadequate. However it may be the case that under certain circumstances an effective potential function of degree $-1$ could emerge for certain subsystems of the universe whose shape is evolving along a geodesic on shape space. Julian Barbour probed this possibility (in \cite{1}) by starting with the following homogeneous function of degree $-2$ \[U=-\frac{W^{2}}{2}\] as the potential on  absolute space, where $W=\sum_{i<j}\frac{m_{i}m_{j}}{\mid x_{i}-x_{j}\mid}$ . Then in an inertial frame of reference with the spacial Newtonian time parameter $t$ (see \eqref{ntime}), the equations of motion are 
\begin{equation}
\frac{dp^{i}}{dt}=-W\frac{\partial W}{\partial x^{i}}
\end{equation}
which is the Newton's law if the $W$ behind the differentiation is replaced with $G$. He then goes on and argues that if the system is virialized,\footnote{virialized means a system of gravitationally interacting particles that is stable. The smaller structures can still interact with each other, but the clusters as a whole doesn't expand or collapse. When a cluster is virialized the merging process and the collapse of matter have finished and the formation process of the galaxy is done. A system is virialized when the potential energy is twice the negative kinetic energy. From this one can find the condition $R_{vir}\cong \frac{R_{max}}{2}$ where $R_{vir}$ is the radius when the cluster is virialized, and $R_{max}$ is the radius (of the moment) at which the cluster starts to collapse. So by looking at the radius and the density of a cluster one can deduce if a cluster is virialized or not.} the value of $W$ remains effectively constant, and hence the motions of Newtonian type (hence described by homogenous potentials of degree $-1$) emerge  effectively.
Note that the vanishing of the Dilational momentum as a consequence of best matching with respect to scale transformations, enforces the constancy of the moment of inertia $I_{cm}$. So one may think that the experimentally verified Hubble expansion is clearly in contradiction with constancy of $I_{cm}$. But as Julian Barbour argued, constancy of $I_{cm}$ would not prevent the matter from clumping, which would increase $-W$ and hence the gravitational constant. Take for instance, a planet orbiting a sun in a universe that in general is becoming clumpier. According to  Barbour's theory, the gravitational force should then become stronger, and hence distance between the planet and the sun should adiabatically decrease. However if one insists that the strength of gravity in the distinguished inertial frame is constant (as is the case in normal Newtonian theory) one is forced to have an adiabatic increase of all scales in this frame. This would then mimic a Hubble-type expansion in this frame. However, he does mention that he probably cannot go around the Hubble red shift in this way... .\newline\newline
Barbour then tried another option to achieve a scale invariant theory. This time by using the unique conserved quantity $I_{cm}$ in his 2003 theory. Newtonian potentials can be converted into scale invariant potentials if multiplied by an appropriate power of $\mu=\sqrt{I_{cm}}=\sqrt{\sum_{i<j}m_{i}m_{j}r_{ij}^{2}}$.\newline
Thus, starting with the general Newtonian potential \[V=\sum_{k=-\infty}^{\infty}a_{k}V_{k}\] with $V_{k}$'s being homogeneous functions of degree $k$,  one goes over to the scale invariant potential \[\tilde{V}= \sum_{k=-\infty}^{\infty} b_{k}V_{k}\mu^{-(2+k)}\] and according to Newton's law of motion, one ends up with the following equations of motion
\begin{equation}\label{ris}
\frac{dp^{i}}{dt}=-\sum_{k=-\infty}^{\infty} b_{k}\mu^{-(2+k)}\frac{\partial V_{k}}{\partial x^{i}}
\end{equation}
\[
+\sum_{k=-\infty}^{\infty} (2+k)b_{k}\mu^{-(2+k)}V_{k}\frac{1}{\mu}\frac{\partial \mu}{\partial x^{i}}\]
In order to make the connection with the observations (hence also to $V$), one has to set $b_{k}\mu^{-(2+k)}=a_{k}$.  Then the equations of motion for the modified (scale-invariant) potential \eqref{ris} turns into the following
\begin{equation}
\frac{dp^{i}}{dt}=-\sum_{k=-\infty}^{\infty}a_{k}\frac{\partial V_{k}}{\partial x^{i}}+C(t)\sum_{j}m_{i}m_{j}\frac{\partial r_{ij}^{2}}{\partial x^{i}}
\end{equation} 
with $C(t)=\frac{\sum_{k=-\infty}^{\infty}(2+k)a_{k}V_{k}}{2\sum_{i<j}m_{i}m_{j}r_{ij}^{2}}$.\newline
So the real force acting on the $i$'th particle can be decomposed into the Newtonian-type forces and a residue which is a cosmological force. However in \cite{17} it is claimed that also this attempt turns out to be empirically inadequate, since it fails to show the formation of clusters for an expanding N-body solutions, and therefore fails to explain the emergence of stars and galaxies.\newline
In a recent publication \cite{2}, these ideas of emergence of interacting theories on absolute space, from a free (non-interacting) theory on Shape space are being elaborated, and expanded to the quantum mechanics. \newline\newline 
In this paper we did not insist on geodesity of dynamics on shape space. We rather sought the projected dynamical law from the absolute configuration space down to the shape space. The principle of relationalism was used to make classical physics invariant under global scale transformations.           

\subsection{Mass tensor}
Originally the Kinetic energy $K$ of a classical $N$-particle system is expressed(defined) in Cartesian coordinates as following
\begin{equation}\label{KEN}
K=\frac{1}{2}\sum_{i=1}^{N}m_{i}\dot{\textbf{r}}_{i}^{2}=\frac{1}{2}[\dot{\textbf{r}}_{1},...,\dot{\textbf{r}}_{N}] \textbf{M} \left[\begin{array}{c}\dot{\textbf{r}}_{1}\\ ...\\ \dot{\textbf{r}}_{N}\end{array}\right]
\end{equation} where $\dot{\textbf{r}}_{i}:=\frac{d\textbf{r}_{i}}{dt}$ with $\textbf{r}_{i}=\left[\begin{array}{c}x_{3i-2}\\ x_{3i-1}\\ x_{3i}\end{array}\right]$ and $\textbf{M}$ is the so called mass matrix which is in this case just a block diagonal $3N\times 3N$ matrix with $\begin{bmatrix}m_{j} & 0 & 0 \\0 & m_{j} & 0\\ 0 & 0 & m_{j} \end{bmatrix}$ as its $j$'s block. 
\newline 
Here, as usual, configuration space is coordinatized by $x_{1},x_{2},...,x_{3N}$ which are in turn the collection of Cartesian coordinates $x_{3i-2}, x_{3i-1}, x_{3i}$ used to denote the position(vector in $\mathbb{R}^{3}$) of the $i$'th particle $\textbf{r}_{i}$.\newline 
Now if the system suffers a number of holonomic constraints, generalized coordinates $q_{1},q_{2},...,q_{f}$ (with $f<3N$ standing for total number of remaining degrees of freedom), can be used for coordinatizing the new (generalized) configuration space.\newline
Now if one rewrites the kinetic energy $K$ in terms of these new generalized coordinates $q_{j}$ and their velocities $\dot{q}_{j}$ one ends up usually with a much more complicated expression than \eqref{KEN} and,  in fact, the metric looses independence of the configuration, and its simple diagonal quadratic form in the velocities. In generalized coordinates, it is quadratic but not necessarily homogeneous in the velocities $\dot{q}_{j}$, and has in general a non-trivial dependence on the coordinates $q_{j}$ (through $M$).\newline
If the coordinate transformation between the set of Cartesian coordinates ${x_{1},...,x_{3N}}$ and the generalized coordinates ${q_{1},...,q_{f}}$ is time-independent (see \cite{10}), the kinetic energy can be written as
\begin{equation}\label{kine}
K=\frac{1}{2}\sum_{k,l}M_{kl}\dot{q}_{k}\dot{q}_{l}=\frac{1}{2}[\dot{q}_{1},...,\dot{q}_{f}]\textbf{M}\left[\begin{array}{c}\dot{q}_{1}\\ ...\\ \dot{q}_{f}\end{array}\right],
\end{equation}  
where \[M_{kl}=\sum_{j=1}^{N}m_{j}\frac{d\textbf{r}_{j}}{dq_{k}}.\frac{d\textbf{r}_{j}}{dq_{l}}\]\[=\sum_{j=1}^{N}\sum_{i=0}^{2}m_{j}\frac{dx_{3j-i}}{dq_{k}}\frac{dx_{3j-i}}{dq_{l}} \] are elements of the $f\times f$ matrix $\textbf{M}$.  \newline
The Lagrangian of classical mechanics is shown to be $L=K-V$, where the potential $V$ is usually independent of the generalized velocities $\dot{q}_{i}$. The conjugate momentum to $q_{i}$ is defined as
\begin{equation}\label{momentum}
p_{i}=\frac{\partial L}{\partial q_{i}}=\frac{\partial K}{\partial \dot{q}_{i}}=\sum_{j=1}^{f}M_{ij}\dot{q}_{j}.
\end{equation} 
Thus the expression \eqref{kine} for the kinetic energy which involved just the velocities can be rewritten as 
\begin{equation}\label{sub}
K=\frac{1}{2}\sum_{i=1}^{f}p_{i}\dot{q}_{i}
\end{equation} 

\subsection{Adjoint and Coadjoint action of a Lie-Group}
Let $G$ be a Lie group, and $\bf G$ its Lie algebra, and $\bf G ^{*}$ be the dual vector space of $\bf G$. The adjoint representation of $g\in G$ on $\bf G$ is defined by
\begin{equation}\label{adjact}
   Ad_{g}(Y)=\frac{d}{dt}\mid_{t=0}(g e^{tY}g^{-1})
\end{equation}
for $Y \in \bf G$.\newline
The coadjoint action of $g\in G$ on $\bf G^{*}$ is characterized by 
\begin{equation}
    <Ad^{*}_{g}(\xi),Y>=<\xi , Ad_{g^{-1}}(Y)>
\end{equation}
for $\xi \in \bf G^{*}$\newline
Here, $<,>:\bf g^{*}\times \bf g \rightarrow R$ is the dual pairing.\newline
In summary: $ad_{g}(x)=gxg^{-1}$, $Ad_{g}=(ad_{g})_{*}:\bf G\rightarrow \bf G$ being called the adjoint action , $Ad^{*}_{g}:\bf G^{*}\rightarrow \bf G^{*}$ being called the coadjoint action. 
\newline
\subsection{Isomorphism $R$}
There exists an isoporphism $R$ between the Lie-algebra $\textbf{so}(3)$ of the rotation group, and the linear space $\wedge^{2}\mathbb{R}^{3}$ of all antisymmetric tensors of order $2$, which we want to explain shortly. \newline
Take ${\pmb{e}_{1},...\pmb{e}_{3}}$ as an orthonormal basis of $\mathbb{R}^{3}$. Then $\pmb{e}_{i}\wedge \pmb{e}_{j}$ with $i<j$ constitutes an orthonormal basis of $\wedge^{2}\mathbb{R}^{3}$.
The inner product in $\wedge^{2}\mathbb{R}^{2}$ is defined as the following
\begin{equation}\label{tvsp}
(u\wedge v \mid x\wedge y)= \begin{vmatrix}
(u\mid x) & (u\mid y) \\ 
(v\mid x) & (v\mid y)
\end{vmatrix}.
\end{equation}
 
One can easily check that for two two-vectors (or tensors of order 2) $\xi=\sum_{i<j}\xi_{ij}\pmb{e}_{i}\wedge \pmb{e}_{j}$ and $\zeta=\sum_{k<l}\zeta_{kl}\pmb{e}_{k}\wedge \pmb{e}_{l}$, definition \eqref{tvsp} leads to the following
\begin{equation}
(\xi\mid\zeta)=\sum_{i<j}\xi_{ij}\zeta_{ij}.
\end{equation}

Now we identify the Lie-algebra of the rotation group in 3 dimensions \textbf{so}(3) with the space of two forms (anti-symmeric tensors) $\wedge^{2}\mathbb{R}^{3}$ by the isomorphism $R$ 
\begin{equation}\label{R}
R: \wedge^{2}\mathbb{R}^{3} \xrightarrow{\sim} \textbf{so}(3)
\end{equation}
\[\xi \rightarrow R_{\xi}.\]
So for $u$,$v$,$x\in \mathbb{R}^{3}$ we define the following
\begin{equation}\label{isoR}
R_{u\wedge v}(x):=(v\mid x)u-(u\mid x)v.
\end{equation}  
$R_{u\wedge v}$ is in fact a 3 dimensional square matrix and its multiplication by a 3 dimensional vector $x$ is given by the last equation. For $\xi \in\wedge^{2}\mathbb{R}^{3}$ and $x=\sum x_{j}\pmb{e}_{j}\in \mathbb{R}^{3}$ one can also write the above formula as 
\begin{equation}
R_{\xi}(x)=\sum_{i}(\sum_{j}\xi_{ij}x_{j})\pmb{e}_{i}.
\end{equation}
That is, $R_{\xi}$ is an antisymmetric matrix with entries $\xi_{ij}$.\newline
Given the natural scalar product of the Lie algebra; $(\alpha\mid\beta)=\frac{1}{2}tr(\alpha\beta^{T})$ for $\alpha$,$\beta \in \textbf{so}$(3) one can show that the identification $R$ is even an isometry from $\wedge^{2}\mathbb{R}^{d}$ to \textbf{so}(d).

As explained in \cite{5}, the space $\wedge^{2}\mathbb{R}^{3}$ can be identified with $\mathbb{R}^{3}$ by $\pmb{e}_{1}\wedge \pmb{e}_{2}\rightarrow \pmb{e}_{3}$ and its cyclic permutations. Hence if one sets \[\xi_{12}=\phi^{3}, \xi_{23}=\phi^{1}, \xi_{31}=\phi^{2}\] the two vector \[\xi=\sum_{i<j}\xi_{ij}\pmb{e}_{i}\wedge \pmb{e}_{j}\] is identified with \[\phi=\sum \phi^{i} \pmb{e}_{i}.\] 
So in this case $R$ becomes a linear isomorphism from $\mathbb{R}^{3}$ to $\textbf{so}(3)$ i.e.
\[R: \mathbb{R}^{3}\rightarrow \textbf{so}(3)\]
\begin{equation}\label{isoR1} 
    R_{\xi}(x)=R_{\phi}(x)=-\phi\times x
\end{equation}
for $x\in\mathbb{R}^{3}$.\newline 
Alternatively, $R_{e_{1}}$ is the matrix $(\xi_{ij})$ with the only nonzero elements $\xi_{23}=-\xi_{32}=1$.\newline   
One can also show(\cite{5}) that $R$ is Ad-equivariant i.e. $R_{g\phi}=Ad_{g}R(\phi)=gR(\phi)g^{-1}$.\newline 
The map $R$, and the inertial tensor $A_{x}$ have the following properties:  
\begin{subequations}\label{properties of R A}
\begin{align}
    R_{a}b=a\times b     \\
    R_{ga}=gR_{a}g^{-1}    \\
    R_{a}.R_{b}=<a\mid b> \\
    A_{gx}(a)=gA_{x}(g^{-1}a):= Ad_{g}A_{x}(a) \\
    (x\mid R_{\xi}y)=(x\wedge y\mid \xi)\\
    (R_{\xi}x\mid R_{\eta}y)=(R_{\xi}x\wedge y\mid \eta)
\end{align}
\end{subequations}
where $a,b\in \mathbb{R}^{3}$ and $g\in SO(3)$. 
\newpage

\subsection{Symbols}

\begin{tabular}{cp{0.3\textwidth}}

     $x$   &  A point on $Q_{cm}:=\frac{Q}{\mathbb{R}^{3}}$ \\
      $q$   &  A point on $Q_{int}=\frac{Q_{cm}}{SO(3)}$ \\
       $s$   &  A point on shape space $S:=\frac{Q}{sim(3)}$ \\
      $dl$   &  line element w.r.t. the mass metric \\
       $\pmb{r}_{i}$  & $i$'s Jacobi vector of an $N$-particle system\\ 
         $\lambda$   & Scale variable of a system\\
         $\dot{\lambda}$   & Scale velocity of a system\\
         $\dot{\pmb{\lambda}}$   & Scale velocity of a system measured in internal units\\
       $\alpha,\beta,\gamma$   & Euler angles connecting a body frame to the space frame\\
         $\{\textbf{e}_{1},\textbf{e}_{2},\textbf{e}_{3}\}$    & Fixed laboratory frame, or space frame\\
         $\{\textbf{e}_{1}',\textbf{e}_{2}',\textbf{e}_{3}'\}$    & Body frame\\
          $\textbf{J}=\sum_{\alpha=1}^{N}m_{\alpha}x_{\alpha}\times\frac{\partial}{\partial x_{\alpha}}$   & Total angular momentum\\
         $\textbf{J}$    & $=\sum_{a=1}^{3}\textbf{e}_{a}J_{a}=\sum_{a=1}^{3}\textbf{e}_{a}'L_{a}$\\
         $J_{a}$   & $=(\textbf{e}_{a}\mid \textbf{J})$\\ 
         $L_{a}$   &  $= (\textbf{e}_{a}'\mid \textbf{J})$\\
          $J_{a}\textbf{r}_{i}$   & $=\textbf{e}_{a}\times \textbf{r}_{i}$\\
           $L_{a}\textbf{r}_{i}$  & $=\textbf{e}_{a}'\times\textbf{r}_{i}=g\big(\textbf{e}_{a}\times\sigma_{i}(q)\big)$\\
          $\omega^{a}(J_{b})$   & $=\delta^{a}_{b}$ \\
          $\omega'^{a}(L_{b})$  & $=\delta^{a}_{b}$\\
          $\pmb{k}$   & Curvature tensor of shape space\\
          $\pmb{c}$   &  Speed of light\\
          $c$   & letter used to characterize scale transformations by a factor $c\in\mathbb{R}^{+}$\\ 
          $\textbf{D}$   &  Dilational tensor\\
          $D$   &  value of system's dilational momentum measured in internal units\\
          $\textbf{M}$   &  Mass metric on $Q_{cm}$ or $Q$\\
          $K$   &  kinetic metric on $Q_{cm}$ or $Q$, i.e. $K=\frac{1}{2}\textbf{M}$\\
          $B$   &  metric on $Q_{int}$\\
         $N$   &  metric on shape space $S$\\
          $A$   &  Moment of inertia tensor\\
           $I_{x}$   &  The canonical isomorphism from tangent space to cotangent space of $Q$ \\
 
   \end{tabular}

\newpage
\textbf{Acknowledgement:} We are greatly thankful to Prof. Dr. Detlef D\"urr for his supervision, generous support, and numerous enlightening discussions during the supervision of one of the authors master thesis, and Ph.D thesis. We are also thankful to Dr. Paula Reichert, Prof. Julian Barbour, Florian Hoffmann, and Dr. Ward Struyven for numerous interesting discussions. We also want to thank Prof. Dr. Christian Maes, for his support for the conduction of one of the author's master thesis.    


\medskip

\bibliographystyle{unsrt}
\bibliography{main}

\begin{thebibliography}{10}

\bibitem{5}
Toshihiro Iwai.
\newblock A geometric setting for classical molecular dynamics.
\newblock 47(2):199--219, 1987.

\bibitem{6}
Jerrold Marsden and Alan Weinstein.
\newblock Reduction of symplectic manifolds with symmetry.
\newblock {\em Reports on mathematical physics}, 5(1):121--130, 1974.

\bibitem{15}
Isaac Newton.
\newblock The principia: mathematical principles of natural philosophy.
\newblock {\em Univ of California Press}, 1999.

\bibitem{16}
Isaac Newton.
\newblock Scholium to the definitions in philosophiae naturalis principia
  mathematica, bk. 1 (1689), trans. andrew motte (1729), rev. florian cajori,
  berkeley: University of california press, 1934. pp. 6-12, paragraph xiv.

\bibitem{19}
Henry~Gavin Alexander.
\newblock The leibniz-clarke correspondence.
\newblock {\em Philosophy}, 32(123), 1956.

\bibitem{14}
Ernst Mach.
\newblock The science of mechanics: A critical and historical exposition of its
  principles.
\newblock {\em Open court publishing Company}, 1893.

\bibitem{13}
Pierre-Louis~Moreau de~Maupertuis.
\newblock Les loix du mouvement et du repos deduite d'un principe metaphysique.
\newblock {\em Histoire de l'Academie Royale des Sciences et des Belles-Lettres
  da Berlin […] pour l'anncee 1746}, 286, 1748.

\bibitem{8}
Chris~G Gray.
\newblock Principle of least action.
\newblock {\em Scholarpedia}, 4(12):8291, 2009.

\bibitem{12}
Cornelius Lanczos.
\newblock The variational principles of mechanics.
\newblock {\em University of Toronto press}, 2020.

\bibitem{11}
Herbert Goldstein.
\newblock Classical mechanics (3rd ed.).
\newblock {\em United States of America: Addison Wesley, 1980}.

\bibitem{1}
Julian Barbour.
\newblock Scale-invariant gravity: particle dynamics.
\newblock {\em Classical and quantum gravity}, 20(8):1543, 2003.

\bibitem{18}
Julian Barbour.
\newblock The nature of time.
\newblock {\em arXiv preprint arXiv:0903.3489}, 2009.

\bibitem{3}
John~D. Barrow.
\newblock The constants of nature: from alpha to omega-the numbers that encode
  the deepest secrets of the universe.
\newblock {\em PANTHEON BOOKS, NEW YORK, 2002}.

\bibitem{4}
Ilse Rosenthal-Schneider.
\newblock Reality and scientific truth: Discussions with einstein, von laue,
  and planck.
\newblock {\em Wayne State University Press, 1980}.

\bibitem{9}
Baptiste Coquinot, Pau~Mir Garcia, and Eva~Miranda Galcer{\'a}n.
\newblock The b-geometry of magnetic fields.
\newblock 2020.

\bibitem{20}
Alan Weinstein.
\newblock A universal phase space for particles in yang-mills fields.
\newblock {\em Letters in Mathematical Physics}, 2(5):417--420, 1978.

\bibitem{21}
Victor Guillemin and Shlomo Sternberg.
\newblock Symplectic techniques in physics.
\newblock {\em Cambridge university press}, 1990.

\bibitem{10}
Robert~G Littlejohn and Matthias Reinsch.
\newblock Gauge fields in the separation of rotations andinternal motions in
  the n-body problem.
\newblock {\em Reviews of modern physics}, 69(1):213, 1997.

\bibitem{2}
Detlef D{\"u}rr, Sheldon Goldstein, and Nino Zangh{\`\i}.
\newblock Quantum motion on shape space and the gauge dependent emergence of
  dynamics and probability in absolute space and time.
\newblock {\em Journal of Statistical Physics}, 180(1):92--134, 2020.

\bibitem{17}
Julian Barbour, Tim Koslowski, and Flavio Mercati.
\newblock A gravitational origin of the arrows of time.
\newblock {\em arXiv preprint arXiv:1310.5167}, 2013.

\end{thebibliography}
\end{document}